\documentclass{JFM-FLM_Au}
\usepackage{todonotes}
\usepackage{xcolor}
\usepackage{gensymb}
\usepackage{pdflscape}

\usepackage[figuresright]{rotating}

\newcommand{\rev}[1]{\textcolor{black}{#1}}

\lefttitle{M. Mohaghar, A. Bhattacharjee, S. S. Jain and D.R. Webster}
\righttitle{Finger-type convection in double-diffusive instability}

\title{Dynamics of finger-type convection in double-diffusive instability}

\author{Mohammad Mohaghar\aff{1}, Anirban Bhattacharjee\aff{2}, Suhas S. Jain\aff{2} \and Donald R. Webster\aff{1}}

\affiliation{\aff{1}School of Civil and Environmental Engineering, Georgia Institute of Technology, Atlanta, GA, USA
\aff{2}\textit{Flow Physics and Computational Science Lab}, George W. Woodruff School of Mechanical Engineering, Georgia Institute of Technology, Atlanta, GA, USA}

\corresau{Mohammad Mohaghar, \email{mohaghar@gatech.edu}}

\begin{document}
\maketitle

\begin{abstract}
Finger-type convection in double-diffusive instability (DDI) controls mixing and scalar transport in many stratified flows, yet a comprehensive, quantitative, finger-resolved description of \rev{transient growth, transport, and saturation remains limited}. Here, finger-type DDI is \rev{analysed} in a sealed-surface laboratory facility using synchronised planar laser-induced fluorescence (PLIF) and particle image velocimetry (PIV) at fixed thermal contrast $\Delta T=5^\circ$C and three salinity contrasts, $\Delta S=350$, 450, and 550~ppm, complemented by \rev{a high-resolution three-dimensional direct numerical simulation (DNS) using the same fluid properties and imposed scalar contrasts}. A systematic fingertip detection and tracking framework generates ensemble growth curves. The dynamics of fingertip growth follow \rev{three stages:} acceleration, quasi-steady propagation, and decay. \rev{The peak growth rates increase monotonically with $\Delta S$ in the intermediate-growth regime and are closely reproduced by DNS, while the nondimensional fingertip-height histories collapse onto a common trend within the investigated parameter range.} The mixed-material area \rev{initially follows} a common nondimensional trend and subsequently \rev{transitions} to $\Delta S$-dependent interaction and breakdown. Finger-scale measurements reveal the formation of a symmetric vortex ring at the \rev{fingertips} for $\Delta S=450$~ppm, which induces vertically aligned transport. \rev{In contrast, at $\Delta S=550$~ppm, both the experiments and DNS exhibit asymmetric roll-up and zig-zag/lateral-drift behaviour, accompanied by enhanced non-vertical transport.} Finally, \rev{buoyancy-anomaly evolution} links the growth-rate phases to a time-dependent force balance in which increasing buoyancy drives acceleration, shear-induced resistance regulates quasi-steady propagation, and dilution \rev{and top-boundary-influence yield} late-stage \rev{fingertip} deceleration.
\end{abstract}



\section{Introduction}
\label{sec:introduction}

Fluid instabilities driven by the combined action of thermal and compositional gradients have been of sustained interest since it was first recognized that differences in molecular diffusivities can induce buoyancy-driven motion.
Early work on thermohaline circulation was motivated by oceanographic observations indicating that vertical transport and fine-scale structures can arise in stratified fluids whose density depends on more than one scalar quantity \citep{stommel1956ocean,stern1960salt}.
The essential insight is that buoyancy-driven evolution depends not only on the instantaneous fluid density field, but also on the relative diffusion rates of its contributing components.
As a result, double-diffusive instabilities (DDI) arise that have no analogue in convection driven by \rev{a single scalar} and form a broad class of flows collectively known as double-diffusive convection \citep{turner1974double,schmitt1994double,radko2013double}.
These processes play an important role in geophysical and astrophysical environments, including oceans, lakes, planetary interiors, and stellar systems, wherever multiple density stratifying agents coexist with unequal rates of molecular diffusivity \citep{schmitt1981form,kunze2003review,denissenkov2010numerical,leconte2012new,leconte2013layered,brown2013chemical}. \rev{Beyond buoyancy-driven systems, differential diffusion can also produce instability in miscible, viscosity-stratified shear flows, where scalars with unequal diffusivities modify the viscosity distribution and destabilise pressure-driven channel flow \citep{sahu2011linear,govindarajan2014instabilities}. These mechanisms differ from the double-diffusive fingering instability considered here but illustrate the broader role of competing diffusivities in hydrodynamic stability.}

To distinguish between the different dynamical responses that may arise in double-diffusive systems, it is convenient to \rev{characterise} the flow using a set of nondimensional parameters that quantify the competing effects of thermal and compositional buoyancy.
A central quantity is the density ratio, defined as

\begin{equation}
R_\rho = \frac{\alpha(\partial T/\partial z)}{\beta(\partial S/\partial z)},
\end{equation}
which measures the stabilising or destabilising contribution of the temperature gradient $(\partial T/\partial z)$ relative to that of the \rev{salinity} gradient $(\partial \rev{S}/\partial z)$. Here, $\alpha$ and $\beta$ denote the thermal expansion and \rev{haline contraction} coefficients, respectively.



In addition, the thermal and compositional Rayleigh numbers, respectively $Ra_T = g \alpha \Delta T H^3 / (\nu \kappa_T)$ and $Ra_S = g \beta \Delta S H^3 / (\nu \kappa_S)$, quantify the relative importance of buoyancy forces compared with viscous and diffusive effects. These parameters are related through the diffusivity ratio $\tau = \kappa_S / \kappa_T$, such that $R_\rho \approx \rev{(Ra_T/Ra_S)/\tau = Le\,(Ra_T/Ra_S)}$ for weakly varying gradients \citep{veronis1965finite,baines1969thermohaline,stern1969collective}, \rev{where $Le=\kappa_T/\kappa_S=1/\tau$ is the Lewis number.} Here, $\kappa_T$ is the thermal diffusivity, $\kappa_S$ is the compositional (i.e., salt) diffusivity, $\nu$ is the fluid kinematic viscosity, $g$ is the acceleration due to gravity, and $H$ is a characteristic length scale of the layer height.


The nature of the resulting flow depends on the value of $R_\rho$ and on the relative signs of the thermal and compositional gradients.
For the configuration in which warm, salty fluid overlies cold, fresh fluid, overturning convection occurs for $R_\rho < 1$, fingering double-diffusive instability arises for $1 < R_\rho < 1/\tau$, and the system is linearly stable for $R_\rho > 1/\tau$ \citep{schmitt1983characteristics,turner1974double,schmitt1994double,radko2013double}.

The present study focuses on the intermediate regime $1 < R_\rho < 1/\tau$, where finger-type DDI controls scalar transport and promotes layer formation.
This regime is therefore important to explain vertical heat and salt transport and the emergence of thermohaline staircases in stratified environments \citep{turner1965coupled,schmitt1979growth,radko2013double}.

The fingering mechanism can be illustrated by a two-layer configuration in which cold, fresh water underlies warm, salty water, arranged such that the mean density increases with depth and the system is stable to large-scale overturning.
In the absence of motion, the interface between the water masses broadens slowly by molecular diffusion.
Consider a small vertical perturbation of fluid across the interface.
Because heat diffuses much more rapidly than salt in water, a displaced fluid parcel rapidly comes into thermal equilibrium with its new surroundings while retaining much of its salinity anomaly during the same time interval.
A fluid parcel displaced downward from the upper layer therefore becomes denser than its new environment and continues to sink. Similarly, a parcel displaced upward from the lower layer becomes lighter than its surrounding environment and continues to rise.
This dynamic mechanism produces a convective instability even though the background density stratification is in a statically stable state.

The instability mechanism favours relatively fine horizontal scales: temperature must equilibrate efficiently between adjacent up- and down-moving regions, which promotes slender structures, whereas excessively small scales are suppressed by viscous dissipation and by lateral diffusion of salinity.
The competing effects balance to form narrow, vertically elongated convection cells that are characteristic of salt fingers.
Finger growth ultimately saturates, as individual fingers reach a finite vertical extent and become unstable to secondary processes, such as shear-driven instabilities or collective interactions among neighbouring fingers \citep{lambert1972vertical,linden1973structure,taylor1996experiments,pringle2002double}.

The fingers transport the salty fluid downward and the fresh fluid upward, thereby modifying the local stratification once saturated.
The accumulation of salt at the base of the fingering zone and fresh water near its top can render adjacent regions unstable to overturning, leading to mixed layers separated by actively fingering interfaces.
This provides a pathway from finger-scale transport to the emergence of thermohaline staircases \citep{merryfield2000origin,krishnamurti2003double,rosenthal2022staircase,radko2003mechanism}.

While this conceptual characterization illustrates how fingering convection can arise, saturate, and promote layer formation, many aspects of the process depend sensitively on the detailed structure of the flow and scalar fields during its evolution.
In particular, the rates at which fingers transport heat and solute, the manner in which they interact and break down, and the conditions under which they trigger mixed-layer formation are inherently dynamical questions that require investigation.
As a result, a range of laboratory studies have aimed at reproducing fingering convection under controlled conditions and at characterising its transport properties \citep{turner1974double,schmitt1994double,radko2013double,tilgner2024experiments}.

A common limitation across much of the experimental literature is that finger dynamics are most often \rev{characterised} through flow visualisation and/or point-probe measurements, whereas quantitative, field-resolved measurements of \emph{both} the scalar quantities and the velocity field at finger scales remain scarce \citep{stern1969salt,schmitt1979growth,taylor1993anisotropy,krishnamurti2006double,tilgner2024experiments}.
As a result, it remains difficult to determine, in a directly measurable way, how finger kinematics (e.g.\ shear, rotation, intermittency, and secondary instability) control local scalar dissipation and flux, and how these processes aggregate into the net transport that underlies the staircase pattern formation.
A related experimental challenge concerns the control and reproducibility of the initial conditions from which fingering instability develops; many classical configurations rely on sharp interfaces, filling procedures, grid stirring, or transient gradient formation, all of which can introduce uncontrolled velocity perturbations or spatial inhomogeneities that influence the early growth and organisation of fingers \citep{stern1969salt,taylor1993anisotropy,tilgner2024experiments}.

\rev{Computational approaches have also been useful}, and direct numerical simulations have become a central tool for exploring nonlinear fingering dynamics.  Most computational configurations remain idealised relative to laboratory facilities, commonly employing periodic or simplified lateral boundary conditions, idealised stratifications, and parameter choices that are constrained by computational cost \citep{yoshida2003numerical,kimura2007direct,garaud2018double}.
Consequently, direct validation of DNS at the level of \emph{finger-resolved} kinematics and scalar structure has remained limited: comparisons have typically relied on bulk fluxes, mean profiles, or qualitative similarity of finger patterns, rather than stringent, field-level tests against simultaneous, co-located velocity and concentration measurements.
Together, these experimental and computational limitations have left open questions regarding the fidelity of predicted tip dynamics, secondary instabilities, and three-dimensional flow organisation under realistic forcing and boundary conditions. Further, these limitations have obscured the role of early-time finger evolution in setting mixed-material production and in partitioning scalar transport between vertical and horizontal pathways.



In this study, these knowledge gaps are addressed through quantitative, finger-resolved measurements of fingering double-diffusive convection initiated from a well-controlled background state. \rev{Relative to earlier experimental studies relying primarily on flow visualisation, the present work combines simultaneous planar laser-induced fluorescence/particle image velocimetry with automated fingertip tracking. The results enable finger growth, lateral deformation, circulation, scalar dissipation rate, and directional transport to be resolved at the scale of individual transient fingers.} This \rev{experimental} approach provides a direct link between finger-scale dynamics and transport processes, clarifying how small-scale instability mechanisms underpin large-scale mixing and the emergence of mixed layers.

To complement the experimental measurements, \rev{a high-resolution three-dimensional direct numerical simulation (DNS) of fingering double-diffusive convection is compared quantitatively with the experimentally measured finger-growth dynamics}. Double-diffusive instability is governed by the coupled evolution of two scalars with disparate diffusivities, and while the experiments provide quantitative, finger-resolved measurements of velocity and \rev{the normalised solutal scalar field}, direct experimental quantification of the temperature field at comparable spatial and temporal resolution remains challenging. Yet, thermal diffusion plays a central role in setting the buoyancy forcing, finger stability, and saturation. The DNS resolves the fully coupled three-dimensional velocity, salinity, and temperature fields, enabling direct examination of thermal gradients and volumetric flow structures that cannot be inferred from planar experimental diagnostics alone. Together, this combined experimental--computational approach provides a comprehensive interpretation of finger growth, saturation, and transport in fingering double-diffusive convection.

The present study addresses several physical questions that have remained difficult to resolve. In particular, it examines whether the early-time growth of fingers admits a reproducible description across buoyancy forcing, \rev{whether the fingertip-height histories collapse onto a common trend when expressed in appropriate nondimensional variables, and whether the DNS results reproduce the experimentally measured fingertip propagation rates.} The role of salinity contrast in controlling the \rev{magnitude} and timing of transport during salt finger evolution is also investigated. Finally, the manner in which increasing buoyancy forcing modifies the pathway from coherent fingers to more disordered states is explored, including the emergence of lateral excursions, secondary-finger generation, and a systematic shift toward enhanced horizontal transport prior to mixed-layer formation.





\section{Methods}
\label{sec:methods}

Laboratory investigations of fingering double-diffusive convection have historically pursued two complementary goals: to realise a robust configuration in which fingers form reproducibly and to quantify the associated transport of heat and solute.
A canonical starting point is a two-layer arrangement in which warm, salty water overlies cold, fresh water \rev{such that} the density distribution \rev{remains} statically stable \citep{turner1964new,stern1969collective,paliwal1980double1,paliwal1980double2}.
Early heat--salt experiments established practical protocols for preparing such initial conditions without premature mixing, for example, by filling procedures that maintain stable stratification during setup and by gentle stirring to homogenise each layer without destroying the interface \citep{stern1969salt}.
In these ``run-down'' configurations, the experiment necessarily remains transient, since the imposed temperature and salinity contrasts are progressively depleted by diffusion and convection.
Diagnostics in these early studies often combined qualitative observations of finger emergence with bulk estimates of transport, obtained from temperature profiling and from salinity sampling or conductivity measurements performed intermittently during the evolution \citep{stern1969salt,schmitt1979growth}.

To mitigate sensitivity to initial conditions and to access longer observation times, subsequent experiments pursued configurations that replace the sharp interface by approximately uniform background gradients or that maintain imposed gradients in a quasi-steady manner \citep{taylor1989laboratory,tilgner2024experiments}.
Gradient-based ``run-down'' experiments created a finite-thickness transition zone in which fingers form and evolve while the mean gradients relax.
These studies provided important qualitative and semi-quantitative constraints on finger geometry and coherence, including the extent to which individual structures span the full depth of the stratified region and how their aspect ratio varies with density ratio \citep{stern1969salt,taylor1993anisotropy}.
A parallel development was the use of optical methods that make fingers readily visible, particularly shadowgraph imaging in transparent cells, which enabled planform observations and facilitated comparative studies of finger spacing and organisation \citep{shirtcliffe1970observations}.
Nevertheless, in many such experiments the velocity field was inferred primarily from visual tracers, while scalar information was obtained from point measurements or layer-averaged sampling, limiting the ability to directly connect local kinematics to local scalar transport \citep{schmitt1979growth,mcdougall1984flux}.

More recently, experimental systems have been designed to maintain concentration differences over long times and thereby reach statistically stationary finger convection.
A prominent class uses semi-permeable membranes separating a convection cell from well-mixed reservoirs, effectively fixing the scalar concentrations at the boundaries while preventing throughflow \citep{krishnamurti1990double,cooper1997experimental,pringle2002double}.
These facilities enabled accurate flux measurements from reservoir budgets and supported extended observation of finger layers and staircase formation, and they also motivated the adoption of modern velocimetry (e.g., particle tracking or particle image velocimetry) alongside optical methods for inferring concentration fields from refractive-index effects \citep{krishnamurti2006double}.
A second approach employs electrochemical (electrodeposition) cells, in which an imposed current maintains an adverse compositional gradient while a \rev{stabilising} temperature gradient is imposed across the same cell \citep{hage2010high,kellner2014transition}.
In these experiments, the compositional flux can be determined precisely from the electrical current, while velocimetry can be applied to \rev{characterise} finger flow velocities and length scales over wide parameter ranges.





\subsection{Experimental facility}
\label{sec:exp}

\begin{figure}
  \centerline{\includegraphics[width=\textwidth]{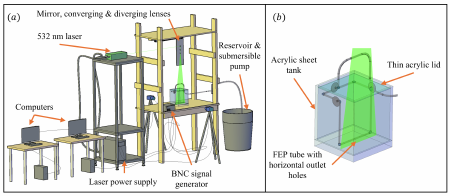}}
  \caption{Schematic of the experimental setup. (\textit{a}) Perspective view showing the overall setup. (\textit{b}) Close-up of the tank and FEP delivery tube with holes for controlled release.}
  \label{fig:setup}
\end{figure}

The objective of the current experiments is to \rev{obtain} high-resolution \rev{simultaneous measurements of the velocity and scalar fields}. Experiments were conducted in a custom-built facility designed for simultaneous planar particle image velocimetry (PIV) and planar laser-induced fluorescence (PLIF) measurements. A schematic of the experimental setup is shown in figure~\ref{fig:setup}. The facility was configured to generate finger-type double-diffusive convection under well-controlled thermal and salinity contrasts. The experiments were performed in a transparent acrylic sheet tank with internal dimensions of $20 \times 20 \times 20~\mathrm{cm}^3$, providing optical access from all sides. The tank was initially filled with hot saline water prepared at the desired background salinity. \rev{The saline solution was prepared by dissolving high-purity food-grade sodium chloride (Morton Culinox 999; $\geq 99.95\%$ NaCl, with no additives) in water in a reservoir and mixing it thoroughly to ensure homogeneity. The solution salinity was measured} with a digital salinity meter. The procedure was repeated for each experiment to achieve the prescribed salinity contrasts.

A thin acrylic lid was placed on top of the tank to create a sealed-surface configuration. This lid suppressed free-surface deformation and evaporation throughout the experiments, ensuring stable boundary conditions at the upper surface.

A fluorinated ethylene propylene (FEP) \rev{delivery} tube was mounted horizontally along the bottom of the tank, as shown in figure~\ref{fig:setup}(b). The tube was perforated with ten identical holes arranged symmetrically, with five holes on each side of the tube. The spacing between adjacent holes was $2~\mathrm{cm}$, resulting in a total \rev{release} span of $8~\mathrm{cm}$ in the horizontal ($x$) direction. The holes were \rev{oriented} such that the \rev{fluid exited horizontally from each hole}\rev{, parallel to the tank bottom}. Cold, fresh water was supplied to the \rev{delivery} tube from an external reservoir by a submersible pump. The flow rate was $3~\mathrm{cc\,min^{-1}}$, such that the characteristic \rev{discharge} velocity from each hole remained below $1~\mathrm{mm\,s^{-1}}$. Under these conditions, the \rev{discharge} introduced negligible vertical momentum, and \rev{because the initial density stratification was statically stable to ordinary overturning convection, the subsequent vertical motion was attributed to double-diffusive instability.}



The refractive index of the FEP \rev{delivery} tube matches that of water, thereby minimizing optical distortion and shadowing in the PIV and PLIF images. This material choice ensured that the surrounding scalar and velocity fields could be measured accurately in the vicinity of the tube.

\subsection{Diagnostics and processing}
\label{sec:diagnostics}

Simultaneous PIV and PLIF measurements were performed to obtain co-located velocity and scalar fields in a vertical plane. Planar illumination was provided by a dual-cavity pulsed Nd:YAG laser emitting at 532~nm, shaped into a thin vertical light sheet using a combination of mirrors and spherical (converging) and cylindrical (diverging) lenses, as shown in figure~\ref{fig:setup}(a). The light sheet was aligned with the central vertical plane of the tank.

Two Oryx 24.6~MP cameras were used to acquire PIV and PLIF images simultaneously for the same field of view. The cameras were arranged on opposite sides of the tank and aligned to view the illuminated plane perpendicularly. Optical filtering was employed to separate the scattered laser light from the tracer particles (used for PIV) from the fluorescence signal used for PLIF, enabling independent and simultaneous recording of the velocity and scalar fields, respectively. For PLIF imaging, Rhodamine~6G dye was mixed with the cold fresh water in the reservoir, and a 550~nm long-pass filter was used on the PLIF camera to block scattered laser light. For the PIV measurements, the flow was seeded with nearly neutrally-buoyant 50~$\mu$m polyamide (Orgasol 2002 D NAT 1, Arkema Inc., King of Prussia, PA) particles (density $\approx 1.02\times10^3~\mathrm{kg\,m^{-3}}$), yielding an estimated particle response time $\tau_p \approx 2\times10^{-4}$~s and a Stokes number $St \sim 10^{-5}$--$10^{-4}$ based on the observed finger length ($L\sim2$--$5$~mm) and velocity ($U\sim0.5$--$1$~mm~s$^{-1}$) scales, indicating negligible particle inertia and faithful flow tracking. The PIV camera was equipped with a 532~nm band-pass filter to block the dye fluorescence wavelengths.

The PIV image pairs were acquired at a sampling rate of 2.5~Hz, corresponding to a fixed time separation of $\Delta t = 0.4$~s between successive frames. For each experimental case, PIV and PLIF images were acquired continuously over a period of 600~s, resulting in a total of 1500 image pairs per case. \rev{Over this acquisition period, the imposed flow rate introduced only $30~\mathrm{cm^3}$ of fresh water, corresponding to less than $1\%$ of the tank volume; therefore, the cumulative inflow did not alter the overall finger-growth behaviour during the 600~s measurement window.} Velocity fields were computed using a multi-pass cross-correlation algorithm with progressively decreasing interrogation window sizes, implemented using DaVis~8.4 (LaVision GmbH). The final interrogation window size was $32\times32$~pixels with 75\% overlap, resulting in a vector spacing of 286~$\mu$m. To reduce small-scale spurious noise, a $5\times5$ Gaussian filter with standard deviation of 0.8 was applied to the velocity fields; the influence of this filtering on small-scale velocity statistics was verified to be negligible. For reference, salt fingers had characteristic widths of approximately 2--5~mm, with tip radii on the order of 4--6~mm, such that each finger was resolved by more than ten velocity vectors across its width.

The PLIF measurements quantified the scalar field associated with salinity-induced buoyancy variations. Raw PLIF images were corrected for background intensity, laser sheet non-uniformity, and attenuation effects prior to further analysis. These corrections were applied consistently across all experiments to ensure quantitative comparability between cases. The processed PLIF fields provided spatially resolved measurements with more than 100~pixels across a typical finger width. Details of the PLIF processing methodology are provided in \citet{mohaghar2019effects} and \citet{mohaghar2023experimental}.



The measured scalar field $C(x,y,t)$ corresponds to the \rev{normalised} concentration of Rhodamine~6G dye added exclusively to the cold, fresh water layer. The dye is dynamically passive and \rev{marks the released fluid, such that $C=1$ represents the cold, fresh-water solution and $C=0$ represents the hot, salty solution}. \rev{Accordingly, $C$ is treated as a normalised solutal scalar. Although Rhodamine~6G and salt do not have identical molecular diffusivities, the salt diffusivity is approximately 3--4 times larger than that of Rhodamine~6G over the temperature range of the present experiments, a difference that is much smaller than the thermal--solutal diffusivity contrast represented by $Le\approx75$. Moreover, their molecular-diffusion lengths over the 600~s acquisition period remain below the characteristic finger width of approximately 2--5~mm. Fluxes of $C$ therefore quantify the resolved finger-scale advective transport of the fresh-water scalar.}

Accurate spatial registration between the PIV and PLIF measurements was achieved using calibration images acquired with both cameras. Dewarped calibration images were aligned using fiducial points and calibration dot patterns, ensuring pixel-to-pixel correspondence between the velocity and scalar fields. This co-registration enabled direct computation of coupled velocity--scalar statistics and facilitated joint visualisation of the velocity, vorticity, and concentration fields. Details of the registration procedure are provided in Appendix~A of \citet{mohaghar2017evaluation}.

\begin{figure}
  \centerline{\includegraphics[width=0.9\textwidth]{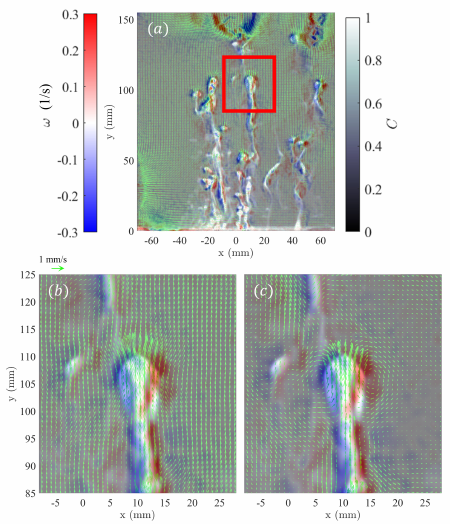}}
  \caption{Example of a velocity--scalar--vorticity field for the case with $\Delta S = 450$~ppm at $t=240$~s. The velocity vectors (green) are overlaid on the vorticity field (red--blue colormap) and the concentration field (grayscale colormap). (\textit{a}) Full field of view showing the spatial distribution of vortical structures and scalar gradients, with the red box indicating the zoom region. (\textit{b}) Zoomed-in view of the selected region, revealing the flow structure around a salt fingertip. (\textit{c}) Same field with the mean velocity $\overline{\boldsymbol{u}} = 0.26$~mm/s (averaged over the zoomed-in region) subtracted from the velocity field, which reveals the relative motion and coherent vortical structures independent of bulk advection.}
  \label{fig:registry}
\end{figure}

Examples of the final registered datasets are shown in figure~\ref{fig:registry} for the case with $\Delta S=450$~ppm. Velocity vectors are overlaid on the corresponding vorticity and concentration fields, illustrating the flow structure around individual salt fingers. The out-of-plane vorticity component is computed from the planar velocity gradients as $\omega=\partial u_y/\partial x-\partial u_x/\partial y$ using central differencing. Figures~\ref{fig:registry}(b,c) present a zoomed-in view of the same region, with figure~\ref{fig:registry}(c) showing the velocity field after subtraction of the local mean velocity. This representation highlights the relative motion and coherent vortical structure near the mushroom-shaped fingertip, independent of bulk advection, and further confirms the fidelity of the combined PIV--PLIF measurements.

\subsection{Experimental cases and governing parameters}
\label{sec:cases}

A series of experiments was conducted to investigate finger-type double-diffusive convection for a fixed thermal contrast and a range of salinity contrasts. In all cases, the imposed temperature difference was held constant at $\Delta T = 5~^\circ\mathrm{C}$, with the hot saline ambient fluid maintained at $40~^\circ\mathrm{C}$ and the cold fresh water at $35~^\circ\mathrm{C}$. The salinity difference was varied as $\Delta S = 350$, $450$, and $550$~ppm. These cases were selected to span a range of density ratios while remaining within the salt-finger regime.

The flow was \rev{characterised} using the thermal and solutal Rayleigh numbers,

\begin{equation}
Ra_T = \frac{g \alpha \Delta T H^3}{\nu \kappa_T}, \qquad
Ra_S = \frac{g \beta \Delta S H^3}{\nu \kappa_S},
\end{equation}
the discrete form of the density ratio,

\begin{equation}
R_\rho = \frac{\alpha \Delta T}{\beta \Delta S},
\end{equation}
where $g$ is the gravitational acceleration, $\alpha$ is the thermal expansion coefficient, $\beta$ is the haline contraction coefficient, $\nu$ is the kinematic viscosity, $\kappa_T$ is the thermal diffusivity, $\kappa_S$ is the salt diffusivity, and $H$ is the \rev{characteristic vertical length scale, taken here as $H=150$~mm}.

The coefficients $\alpha$ and $\beta$ were computed using the Thermodynamic Equation of Seawater 2010 (TEOS--10) at the corresponding temperature and salinity for each experimental condition. \rev{Representative rounded values of the physical properties are listed in table~\ref{tab:physprops}.} For the present experiments, \rev{$Le \approx 75$} indicates a strong contrast between thermal and solutal diffusivities.


The experimental cases and the corresponding nondimensional parameters are summarised in table~\ref{tab:cases}. The thermal Rayleigh number was held fixed at \rev{$Ra_T = 5.9\times10^{8}$} across all cases due to the constant thermal forcing, whereas the solutal Rayleigh number increased with $\Delta S$. As a result, the density ratio decreased from \rev{$R_\rho=7.0$ to $4.4$} as $\Delta S$ increased from 350 to 550~ppm.

\begin{table}
  \centering
  \caption{Physical parameters used for the calculation of nondimensional groups.}
  \label{tab:physprops}
  \begin{tabular}{ll}
    \hline
    Quantity & Value \\
    \hline
    Thermal expansion coefficient & $\alpha = \rev{3.5\times10^{-4}}~^{\circ}\mathrm{C}^{-1}$ \\
    Haline contraction coefficient & $\beta = \rev{7.2\times10^{-4}}~\mathrm{PSU^{-1}}$ \\
    Kinematic viscosity & $\nu = \rev{6.6\times10^{-7}}~\mathrm{m^2\,s^{-1}}$ \\
    Thermal diffusivity & $\kappa_T = \rev{1.5\times10^{-7}}~\mathrm{m^2\,s^{-1}}$ \\
    Salt diffusivity & $\kappa_S = \rev{2.0\times10^{-9}}~\mathrm{m^2\,s^{-1}}$ \\
    Gravitational acceleration & $g = 9.81~\mathrm{m\,s^{-2}}$ \\
    \hline
  \end{tabular}
\end{table}


\begin{table}
  \centering
  \caption{Summary of experimental cases and corresponding nondimensional parameters.}
  \label{tab:cases}
  \begin{tabular}{ccccc}
    \hline
    $\Delta T~(^\circ\mathrm{C})$ & $\Delta S~(\mathrm{ppm})$ &
    $Ra_T$ & $Ra_S$ & $R_\rho$ \\
    \hline
    5 & 350 & \rev{$5.9\times10^{8}$} &
    \rev{$6.4\times10^{9}$} & \rev{7.0} \\
    5 & 450 & \rev{$5.9\times10^{8}$} &
    \rev{$8.2\times10^{9}$} & \rev{5.5} \\
    5 & 550 & \rev{$5.9\times10^{8}$} &
    \rev{$1.0\times10^{10}$} & \rev{4.4} \\
    \hline
  \end{tabular}
\end{table}

\subsection{Automated fingertip detection}
\label{sec:fingerdetection}

\begin{figure}
  \centerline{\includegraphics[width=1.05\textwidth]{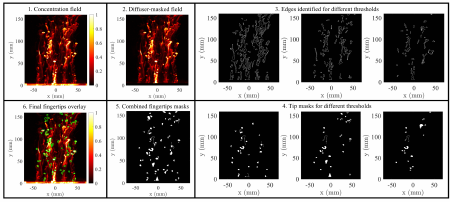}}
  \caption{Example sequential processing steps for automated fingertip detection: (1) concentration field; (2) diffuser region masked; (3) Canny edge detection applied at multiple intensity thresholds (three thresholds shown) to identify fingertip edges; (4) individual tip masks formed from closed-loop contours matching the fingertip geometry; (5) combined fingertip mask created by merging all fingertip masks for varying thresholds; and (6) contours of the fingertip masks overlaid on the original concentration field to \rev{visualise} the spatial distribution of all detected fingertips.}
  \label{fig:tip_pipeline}
\end{figure}

\begin{figure}
  \centerline{\includegraphics[width=\textwidth]{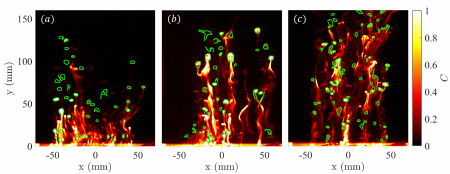}}
  \caption{Demonstration of the fingertip detection method for three salinity contrasts at the same experimental time: (\textit{a}) $\Delta S = 350$~ppm, (\textit{b}) $\Delta S = 450$~ppm, and (\textit{c}) $\Delta S = 550$~ppm. In each case, the detected fingertip contours (green) are overlaid on the corrected concentration field at $t=240$~s, illustrating the method’s capability to identify fingertip locations across a range of salinity contrasts.}
  \label{fig:tip_demo}
\end{figure}

An automated image-based algorithm was developed to identify and track fingertips from the PLIF concentration fields. The method operates directly on the concentration fields and is designed to robustly detect fingertip locations across a range of salinity contrasts and flow intensities. An overview of the sequential processing steps is shown in figure~\ref{fig:tip_pipeline}, and representative detection results are shown in figure~\ref{fig:tip_demo}.

The procedure begins with the concentration field obtained from the calibrated PLIF fields (figure~\ref{fig:tip_pipeline}(1)). Prior to feature extraction, the near-wall diffuser region at the bottom of the tank was masked and removed from the analysis to prevent spurious detections associated with the \rev{delivery} line and enhanced mixing near the source (figure~\ref{fig:tip_pipeline}(2)). This masking was applied consistently to all frames using a fixed vertical cutoff based on the physical location of the diffuser.

Fingertip candidates were identified using an edge-based detection approach applied to the concentration field. Canny edge detection was performed on contrast-adjusted versions of the concentration field over a range of intensity thresholds (figure~\ref{fig:tip_pipeline}(3)). The use of multiple thresholds was found to improve robustness by capturing fingertip outlines across varying local concentration gradients and illumination conditions. For each threshold level, small isolated edge fragments were removed, and morphological closing operations were applied to promote the formation of closed contours.

Closed-loop contours were then \rev{analysed} to identify regions consistent with fingertip geometry (figure~\ref{fig:tip_pipeline}(4)). Candidate regions were required to satisfy constraints on area and eccentricity, thereby excluding elongated filamentary structures, diffuse background features, and large non-fingertip blobs. In cases where contours were not fully closed, convex hull operations were used to form closed regions prior to geometric filtering. Individual fingertip masks were generated for each threshold level and subsequently merged to form a combined fingertip mask (figure~\ref{fig:tip_pipeline}(5)). Additional area-based filtering was applied to the combined mask to remove residual large-scale structures not associated with fingertips. The final detected fingertip contours were overlaid on the original concentration field to \rev{visualise} the spatial distribution of fingertip locations (figure~\ref{fig:tip_pipeline}(6)).

 Figure~\ref{fig:tip_demo} demonstrates the performance of the detection algorithm for three salinity contrasts at the same experimental time. As the salinity contrast increases, a larger number of fingertips are detected, accompanied by a reduction in the characteristic horizontal spacing between neighbouring tips. This trend reflects the enhanced buoyancy forcing at higher salinity contrasts, which leads to increased finger activity and more rapid fingertip evolution over a common experimental time period.

For quantitative analysis, fingertip centroid locations were computed from the final combined mask. Temporal linking of fingertip positions between successive frames was performed using a nearest-neighbour criterion with physically motivated tolerances on horizontal and vertical displacements. This tracking procedure enabled reconstruction of individual fingertip trajectories while allowing for intermittent disappearance and re-emergence due to merging, splitting, or local loss of contrast.


\subsection{Three-dimensional simulation setup}
\label{sec:dns}

Computational studies have provided a complementary route for isolating the mechanisms that govern fingering double-diffusive convection and its secondary processes. Early theoretical and numerical work established the parameter ranges for fingering and diffusive modes and clarified the dependence of growth rates and flux ratios on the density ratio and diffusivity contrast \citep{piacsek:1980,holyer1981collective,shen1989evolution,shen:1995}. With increasing computational capability, direct numerical simulation (DNS) became a primary tool for investigating the fully nonlinear regime, including finger interactions, the emergence of larger-scale organisation, and pathways toward layer formation \citep{merryfield2000origin,rosenblum2011turbulent,medrano2014double,garaud20152d}. More recent DNS studies have systematically explored transport scalings, multiple turbulent states, and flow morphology over broader parameter ranges and in three dimensions, including regimes relevant to high-Rayleigh-number convection and to varying Prandtl/Schmidt-number combinations \citep{yang2016convection,yang2016scaling,yang2020multiple,howland2023double}.

\begin{figure}
  \centerline{\includegraphics[width=\textwidth]{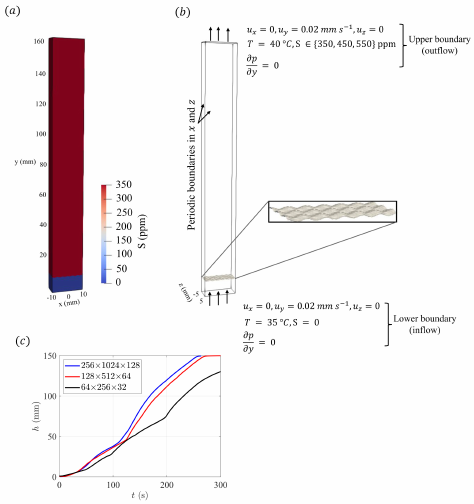}}
  \caption{Three-dimensional simulation configuration, \rev{boundary conditions,} and grid convergence assessment. (\textit{a}) Computational domain showing the initial \rev{salinity} field \rev{for the $\Delta S=350$~ppm case}, with hot, salty fluid (red) overlying cold, fresh fluid (blue). (\textit{b}) Iso-surface of the initial interface perturbation \rev{with the imposed boundary conditions.} (\textit{c}) Grid-sensitivity study for the $\Delta S=450$~ppm case showing the temporal evolution of \rev{fingertip} height $h(t)$ for three grid resolutions: $64 \times 256 \times 32$, $128 \times 512 \times 64$, and $256 \times 1024 \times 128$.}
  \label{fig:cfd_setup}
\end{figure}

The objective of the current simulations is to resolve the three-dimensional velocity, salinity, and temperature fields and to complement the planar experimental diagnostics. \rev{A three-dimensional simulation was performed under the Boussinesq approximation using the same fluid properties and imposed temperature and salinity contrasts as the laboratory cases. Because resolving the full experimental tank volume at the spatial resolution required for DNS would be computationally prohibitive, a reduced subdomain with physical dimensions of $2$~cm, $16$~cm, and $1$~cm in the $x$, $y$, and $z$ directions, respectively, was simulated. Periodic boundary conditions in the horizontal directions were used such that the reduced domain represents an interior portion of the finger field rather than a wall-bounded subdomain. The analysed finger region is sufficiently far from the experimental lateral walls that sidewall effects are not expected to materially influence the finger-scale dynamics examined here; accordingly, the lateral walls are not reproduced explicitly. The discrete diffuser holes are also not reproduced in the numerical domain.} Figure~\ref{fig:cfd_setup} shows the computational configuration.




The governing equations consist of the incompressible continuity equation, Navier–Stokes equations with buoyancy forcing, and advection–diffusion equations for temperature and salinity:

\begin{align}
    \nabla \cdot \textbf{u} &= 0, \label{eq:cont} \\
    \frac{\partial \rho \textbf{u}}{\partial t} + \nabla \cdot (\rho \mathbf u \mathbf u) &= -\nabla p + \nabla \cdot (\mu \nabla \mathbf u) + \rho \textbf{g}(\beta\Delta S - \alpha \Delta T), \label{eq:NS} \\
    \frac{\partial T}{\partial t} + \nabla \cdot (\textbf{u}T) &= \kappa_T \nabla^2 T, \label{eq:transT} \\
    \frac{\partial S}{\partial t} + \nabla \cdot (\textbf{u}S) &= \kappa_S \nabla^2 S, \label{eq:transS}
\end{align}
where $\textbf{u} = \{u_x, u_y, u_z\}$ is the velocity vector, $p$ is the pressure, $\rho$ is the reference fluid density, $\mu$ is the dynamic viscosity, $\beta$ and $\alpha$ are the haline contraction and thermal expansion coefficients, respectively, and $\textbf{g} = \{0,-9.81 \,\mathrm{m\,s^{-2}},0\}$ is gravity. The scalars $T$ and $S$ denote temperature and salinity, with diffusivities $\kappa_T$ and $\kappa_S$, respectively.

Hot, salty fluid ($T = 40^\circ$C, $S = 350$, $450$, or $550$~ppm) was initially placed above cold, fresh fluid ($T = 35^\circ$C, $S = 0$~ppm). \rev{To reduce the computational cost associated with the long initial diffusive transient before appreciable double-diffusive finger growth developed, a weak spatially uniform upward throughflow of $u_y=0.02~\mathrm{mm\,s^{-1}}$ was imposed. The same velocity was prescribed at the lower and upper boundaries, corresponding to inflow at the bottom and equal outflow at the top, so that the net volumetric flux through the domain is zero. Accordingly, the upper boundary is not treated as a no-slip wall; it permits the imposed outflow. This small imposed velocity substantially shortens the initial transient while remaining much smaller than the subsequent double-diffusive fingertip propagation velocities. It is not intended to reproduce the discrete, horizontally directed discharge from the experimental diffuser. Temperature and salinity were held at the corresponding cold, fresh and hot, salty solution compositions at the lower and upper boundaries, respectively. The imposed throughflow velocity was subtracted when evaluating the DNS fingertip propagation rates. Pressure was periodic in the horizontal directions and satisfied a Neumann condition at the vertical boundaries. The complete boundary conditions are summarised in figure~\ref{fig:cfd_setup}.}



The initial interface between the two layers was located at $y = 10$ mm and \rev{was} perturbed according to

\begin{equation}
    y(x,z) = 10 + 0.2 \cos(0.6\pi x) + 0.2 \cos(0.6\pi z).
\end{equation}


This perturbation introduced crests and valleys (figure~\ref{fig:cfd_setup}b) that subsequently developed into evolving salt fingers. The imposed horizontal wavenumber was selected to be consistent with the experimentally observed initial finger spacing. Specifically, application of the automated fingertip detection algorithm to the PLIF concentration fields showed that within the first centimeter above the \rev{delivery} line, approximately three fingertips per centimeter were present during the early growth stage.



The simulations were performed using the finite-volume solver ExaFlow3D \citep{jain2022accurate,jain2024model,jain2025stationary,jain2025model}. Spatial derivatives were discretized using second-order central differences on a staggered grid to preserve kinetic energy consistency, and time integration was performed using a fourth-order Runge–Kutta scheme. For the baseline mesh, the grid spacings correspond to $\Delta x = \Delta z = 0.0078125$ cm and $\Delta y = 0.015625$ cm, with a time step of $\Delta t = 0.002$ s to ensure stability.

Grid convergence was assessed by comparing three resolutions: $64 \times 256 \times 32$, $128 \times 512 \times 64$, and $256 \times 1024 \times 128$. Figure~\ref{fig:cfd_setup}(c) shows the temporal evolution of the \rev{fingertip} height $h(t)$ for the three mesh resolutions in the $450$ ppm case. The coarsest grid underpredicts the vertical growth rate throughout the evolution and exhibits delayed saturation relative to the finer grids. In contrast, simulations for the intermediate and finest grid resolutions show close agreement over the entire growth period. A quantitative error assessment relative to the finest grid resolution results indicates that the results of the intermediate grid resolution exhibit an RMS deviation of 3.2\%, with a maximum deviation of 5.5\%. In contrast, the results for the coarsest grid show significantly larger discrepancies, with an RMS deviation of $18.4\%$ and a maximum deviation of $31.9\%$. Thus, refinement from the coarsest to the intermediate grid reduces the error by approximately a factor of six, indicating systematic convergence of the solution under grid refinement.

The relatively small discrepancy between the intermediate and finest grids, together with the substantial error reduction under refinement, indicates that the solution is approaching grid convergence for the intermediate grid resolution and is effectively grid-independent at the finest resolution for the purpose of capturing global finger growth dynamics. The finest grid resolution was therefore adopted for all \rev{analyses presented here}.


\section{Results}
\label{sec:resultsI}

\subsection{Growth rate}
\label{sec:growthrate}

\subsubsection{Experimental growth rate}
\label{sec:growthrate_exp}

\begin{figure}
  \centerline{\includegraphics[width=\textwidth]{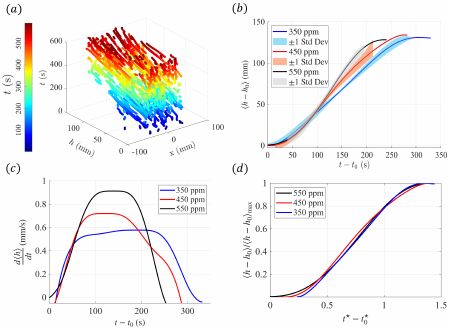}}
  \caption{Experimental finger growth analysis across salinity contrasts: (\textit{a}) trajectories of all detected fingertips for the $\Delta S = 450$~ppm case (as an example), where $h$ is the finger height and the trajectories are additionally colored as a function of time-of-initiation. (\textit{b}) Ensemble-averaged fingertip height $\langle h-h_0\rangle$ versus time $(t-t_0)$ for $\Delta S = 350$, 450, and 550~ppm, with shaded $\pm 1$ standard deviation bands; $t_0$ and $h_0$ are the first-detection time and height, respectively, for each tip. (\textit{c}) Tip growth rate. (\textit{d}) \rev{Nondimensionalised} time record of finger height.}
  \label{fig:growth}
\end{figure}

The temporal evolution of fingertip growth for each salinity contrast from the experimental results is summarised in figure~\ref{fig:growth}. The analysis is based on the automated tracking of individual fingertips extracted from the PLIF fields, as described in \S\ref{sec:fingerdetection}.


Figure~\ref{fig:growth}(a) presents stacked three-dimensional trajectories of all detected fingertips for the case $\Delta S = 450$~ppm. Each \rev{trajectory} represents the vertical evolution of a single fingertip as a function of time. \rev{Because cold, fresh water is supplied continuously through the delivery tube near the tank bottom, double-diffusive fingers initiate throughout each experiment,} resulting in a sequence of fingertip trajectories rather than a single generation event.


For quantitative comparison, each tracked trajectory $i$ is shifted using \rev{the acquisition time $t_{0,i}$ of its first detected frame and the corresponding fingertip-centroid height $h_{0,i}=h_i(t_{0,i})$}, and \rev{the} shifted variables


\begin{equation}
\tilde{t} = t - t_{0,i},
\qquad
\tilde{h}_i(\tilde{t}) = h_i(t) - h_{0,i}
\end{equation}

\noindent
are constructed. More than one hundred \rev{shifted} fingertip trajectories are included in the ensemble for each salinity contrast to obtain a single representative growth curve per case,

\begin{equation}
\langle \tilde{h} \rangle (\tilde{t})
= \frac{1}{N(\tilde{t})}
\sum_{i=1}^{N(\tilde{t})}
\tilde{h}_i(\tilde{t}),
\end{equation}

\noindent
where $N(\tilde{t})$ denotes the number of trajectories contributing at shifted time $\tilde{t}$.

The ensemble-averaged height histories are shown in figure~\ref{fig:growth}(b), and the growth rates are presented in figure~\ref{fig:growth}(c). All three salinity contrasts exhibit a common temporal structure \rev{characterised} by three distinct phases.

In the early stage, newly formed fingertips accelerate upward as local buoyancy forces intensify. During this phase, the growth rate increases with time, reflecting the progressive establishment of buoyancy-driven convection.

In the intermediate stage, finger growth proceeds approximately linearly in time, corresponding to a near-constant growth rate. This regime coincides with the strongest convective transport. As $\Delta S$ increases, the buoyancy forcing becomes stronger, resulting in a larger peak growth rate and a shorter time required for fingers to reach the upper wall.

In the late stage, fingertips approach and interact with the top wall and neighbouring structures. The effective buoyancy contrast weakens, lateral interactions intensify, and the growth rate decreases toward zero. The decay phase occurs earlier for larger $\Delta S$, consistent with the earlier arrival of fingers at the upper wall.

To assess similarity across buoyancy forcing, the ensemble-averaged growth histories are expressed in nondimensional form. A characteristic time scale is defined as

\begin{equation}
t_c = \frac{L_c}{\dot{h}_{\max}},
\qquad
t^{*} = \frac{t}{t_c},
\label{eq:tstar_def_results}
\end{equation}

\noindent
where $L_c$ denotes the \rev{characteristic vertical length scale}, taken here as the total vertical extent of the domain ($L_c = 150$~mm), and $\dot{h}_{\max}$ represents the maximum ensemble-averaged growth rate obtained from figure~\ref{fig:growth}(c). The values of $\dot{h}_{\max}$ are $0.55$, $0.69$, and $0.89$~mm~s$^{-1}$ for $\Delta S = 350$, $450$, and $550$~ppm, respectively. Consistent with the temporal shifting applied prior to ensemble averaging, the ensemble-averaged growth curves are reported as functions of $(t^{*}-t_0^{*})$\rev{, where $t_0^{*}$ denotes the nondimensional first-detection time}.




The resulting nondimensional growth curves are shown in figure~\ref{fig:growth}(d). A close collapse of behaviour for the three salinity-contrast cases is observed over the full evolution. This behaviour indicates that, within the present parameter range, increasing $\Delta S$ primarily rescales the characteristic growth time and height, while the functional form of the three-stage growth dynamics remains approximately invariant.

\subsubsection{Computational growth rate and validation}
\label{sec:growthrate_dns}

\begin{figure}
  \centerline{\includegraphics[width=\textwidth]{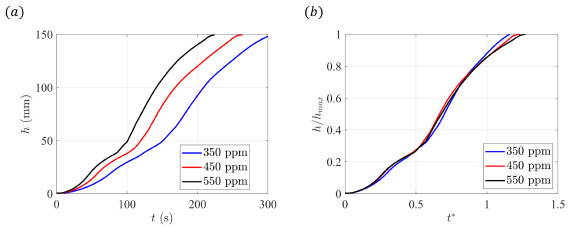}}
  \caption{Computational finger growth analysis across salinity contrasts: (\textit{a}) time evolution of finger height $h$ versus time $t$ for salinity contrasts of 350, 450, and 550~ppm. (\textit{b}) \rev{Nondimensionalised} representation of the same data (finger height is \rev{normalised} by the maximum height $h_{\max}$ and time is scaled by the characteristic growth timescale $t_c$).}
  \label{fig:compGrowth}
\end{figure}

Figure~\ref{fig:compGrowth} shows the corresponding fingertip growth dynamics obtained from the simulation. \rev{In the experiments, double-diffusive fingers initiate at different times throughout the run while cold, fresh water is supplied continuously through the delivery tube, whereas the DNS initializes a one-time finger population simultaneously from the perturbed interface. No additional finger population is introduced after the initial onset; therefore, the computed trajectories represent the collective evolution of one initially generated population rather than the continuously replenished experimental ensemble.}

The time evolution of the computed finger height is shown in figure~\ref{fig:compGrowth}(a) for $\Delta S = 350$, 450, and 550~ppm. As in the experimental results, faster growth and earlier arrival at the upper boundary are obtained as $\Delta S$ is increased. \rev{The nondimensionalised curves also follow a common trend across the three simulated cases, comparable to the experimental result in figure~\ref{fig:growth}(d).}



A subtle difference from the experimental ensemble-mean curves is indicated by modest variations in the slope during the intermediate-growth regime in figure~\ref{fig:compGrowth}(a). \rev{Comparable small variations are also observed in individual experimental fingertip trajectories in figure~\ref{fig:growth}(a), whereas averaging more than 100 shifted trajectories smooths these temporal and finger-to-finger variations in figure~\ref{fig:growth}(b). The variations are small relative to the overall intermediate-stage growth and do not substantially affect the propagation rate averaged over this interval.} As the simulated fingers evolve, progressive thinning of the finger cores and sharpening of scalar gradients occur, which is accompanied by a modest increase in the slope of $h(t)$. When averaged over the full intermediate-growth interval, however, the effective mean growth rates remain consistent with the experimental values.

Specifically, the mean intermediate-regime growth rates obtained from the simulation are $0.57$, $0.70$, and $0.85~\mathrm{mm\,s^{-1}}$ for $\Delta S=350$, $450$, and $550$~ppm, respectively. \rev{As summarized in table~\ref{tab:hdot_comparison}, these values closely agree with the corresponding experimental peak/plateau growth rates of $0.55$, $0.69$, and $0.89~\mathrm{mm\,s^{-1}}$. The DNS is therefore used in subsequent sections to examine three-dimensional flow mechanisms and thermal-field evolution that cannot be resolved by the planar measurements.}


\begin{table}
\centering
\caption{\rev{Comparison of experimentally measured and DNS-derived
intermediate-stage fingertip propagation velocities for varying
salinity contrast.}}
\label{tab:hdot_comparison}
\begin{tabular}{c c c}
\hline
$\Delta S$ (ppm) &
$\dot{h}_{\mathrm{exp}}$ (mm\,s$^{-1}$) &
$\dot{h}_{\mathrm{DNS}}$ (mm\,s$^{-1}$) \\
\hline
350 & 0.55 & 0.57 \\
450 & 0.69 & 0.70 \\
550 & 0.89 & 0.85 \\
\hline
\end{tabular}
\end{table}

\subsection{Mixed-material area, dissipation rate, and scalar-flux diagnostics}
\label{sec:mixing_transport}

The consequences of finger growth for mixing and scalar transport are quantified using (i) a mixed-material metric formed from the PLIF concentration field, (ii) scalar-dissipation-rate and flux maps, and (iii) spatially averaged transport measures. Throughout this section, $C(x,y,t)$ denotes the dimensionless \rev{normalised solutal scalar field} measured by PLIF in the illuminated plane, $\boldsymbol{u}=(u_x,u_y)$ is the planar velocity field obtained from PIV, $A$ denotes the measurement-domain area, and an overbar denotes the spatial average over $A$.





The mixed-material area is defined as

\begin{equation}
A_{\mathcal{M}}(t)
\equiv
\int_A 4\,C(1-C)\,\mathrm{d}A,
\label{eq:Am_def_results}
\end{equation}

\noindent
\rev{where the local weighting $4C(1-C)$ vanishes in unmixed regions ($C=0$ or $1$) and reaches its maximum at $C=0.5$ \citep{mohaghar2017evaluation,mohaghar2019transition}. Thus, $A_{\mathcal{M}}$ quantifies the planar extent of partially mixed material, while the scalar dissipation rate defined below provides a complementary gradient-based measure.}

The \rev{solutal-scalar dissipation-rate estimate} is defined as

\begin{equation}
\chi(x,y,t)
=
\kappa_S\,\lvert \boldsymbol\nabla C \rvert^{2},
\label{eq:chi_def_results}
\end{equation}

\noindent
where $\boldsymbol\nabla C$ is evaluated in the measurement plane.


The planar \rev{solutal-scalar flux} vector is defined as

\begin{equation}
\boldsymbol{\mathcal{J}}_{S}(x,y,t)
\equiv
C(x,y,t)\,\boldsymbol{u}(x,y,t),
\label{eq:Js_plane_def}
\end{equation}
such that its horizontal and vertical components, respectively, are

\begin{equation}
\mathcal{J}_{S,x}=C u_x,
\qquad
\mathcal{J}_{S,y}=C u_y.
\end{equation}


\rev{To quantify the overall intensity of planar advective scalar transport without cancellation between oppositely directed contributions, the magnitude of the flux vector is defined as}

\begin{equation}
\mathcal{J}(x,y,t)
=
\left|\boldsymbol{\mathcal{J}}_{S}\right|
=
C\sqrt{u_x^2+u_y^2}.
\label{eq:Jmag_def}
\end{equation}
\rev{and the corresponding area-averaged transport magnitude is defined as}

\begin{equation}
\overline{\mathcal{J}}(t)
=
\frac{1}{A}
\int_A
C\,\sqrt{u_x^2+u_y^2}
\,\mathrm{d}A.
\label{eq:Jbar_def}
\end{equation}


In the field maps, the divergence of the planar \rev{solutal-scalar flux} is reported as

\begin{equation}
\boldsymbol\nabla\!\cdot\boldsymbol{\mathcal{J}}_{S}
=
\frac{\partial}{\partial x}(C u_x)
+
\frac{\partial}{\partial y}(C u_y),
\label{eq:divJs_def}
\end{equation}
which identifies regions of local \rev{scalar-flux} convergence and divergence within the measurement plane.

\rev{For the numerical simulations, the corresponding component-wise scalar-transport magnitudes are evaluated in three dimensions.} The \rev{normalised solutal and thermal scalar fields} are denoted by $C_S(\boldsymbol{x},t)$ and $C_T(\boldsymbol{x},t)$, respectively, and the velocity field is $\boldsymbol{u}=(u_x,u_y,u_z)$. \rev{Both scalar fields use the same normalisation convention as the experimental field $C$, with $C_S=1$ and $C_T=1$ in the cold, fresh fluid, and $C_S=0$ and $C_T=0$ in the hot, salty fluid.} The \rev{solutal-scalar and thermal-scalar flux vectors} are defined as $\boldsymbol{\mathcal{J}}_{S}=C_S\boldsymbol{u}$ and $\boldsymbol{\mathcal{J}}_{T}=C_T\boldsymbol{u}$. Spatial averages are then computed over the full computational domain of volume $V$, such that


\begin{equation}
\overline{\mathcal{J}}_{S,i}(t)=\frac{1}{V}\int_V \left| C_S u_i \right| \,\mathrm{d}V,
\qquad
\overline{\mathcal{J}}_{T,i}(t)=\frac{1}{V}\int_V \left| C_T u_i \right| \,\mathrm{d}V,
\end{equation}

\noindent
where $i\in\{x,y,z\}$. Thus, the experimental transport measures represent planar area averages, whereas the DNS results correspond to full three-dimensional volume averages of the magnitudes of the scalar-flux components and additionally provide access to \rev{thermal-scalar flux components}.

\subsubsection{Experimental evolution of mixed-material area and \rev{scalar transport magnitude}}
\label{sec:mix_transport_exp}

\begin{figure}
  \centerline{\includegraphics[width=\textwidth]{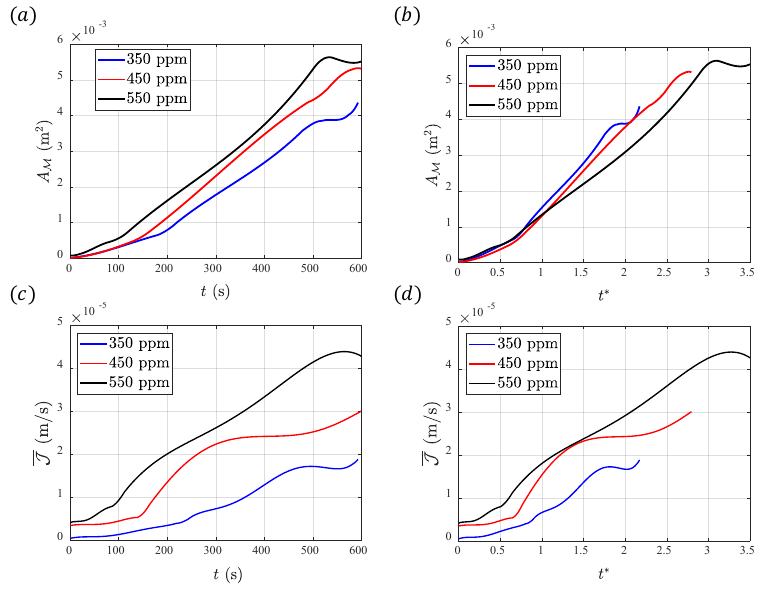}}
  \caption{Temporal evolution of mixed-material area and \rev{planar scalar transport magnitude} for the three experimental salinity contrasts. (\textit{a,b}) Mixed-material area. (\textit{c,d}) \rev{Area-averaged planar scalar transport magnitude}.}
  \label{fig:mix_transport}
\end{figure}

\rev{Figure~\ref{fig:mix_transport} shows the temporal evolution of $A_{\mathcal{M}}$ and $\overline{\mathcal{J}}$ for the three experimental salinity contrasts.} In the early stage, a slow and approximately gradual increase of both $A_{\mathcal{M}}$ and $\overline{\mathcal{J}}$ is observed, consistent with the onset and acceleration stage of fingertip growth (figure~\ref{fig:growth}c), during which the interfacial region broadens and coherent fingers begin to form. At later times, a clear increase in slope is observed in both quantities, indicating the onset of a rapid-mixing/entrainment phase in which lateral interaction, merging, and relaunch events become more frequent and the interfacial area occupied by mixed fluid expands more rapidly.

A systematic dependence on salinity contrast is observed in the timing of this transition (figure~\ref{fig:mix_transport}a). For smaller $\Delta S$, delayed onset of the rapid-mixing phase is obtained, consistent with weaker buoyancy forcing and slower finger growth. For larger $\Delta S$, the rapid-mixing phase occurs earlier, consistent with earlier arrival of fingers to greater heights and earlier interaction with neighbouring structures and the upper boundary.

When time is expressed in nondimensional form (figure~\ref{fig:mix_transport}b), behaviour collapse is obtained during the early growth/entrainment stage, indicating that the initial production of mixed material follows a dynamically similar pathway across the three cases once the characteristic growth time scale is accounted for. After the mixing transition, the collapse is no longer maintained, and the curves diverge. A modestly larger $A_{\mathcal{M}}$ is obtained for the lower-$\Delta S$ cases at late times, consistent with the longer physical time available for lateral interactions and merging prior to strong top-wall influence, whereas the higher-$\Delta S$ case reaches comparable heights more rapidly, and thus the fingers spend less time in the interaction-dominated regime before growth is arrested.


\rev{The area-averaged planar scalar transport magnitude $\overline{\mathcal{J}}$ increases strongly with $\Delta S$ in both dimensional and nondimensional time (figure~\ref{fig:mix_transport}c,d). In contrast to $A_{\mathcal{M}}$, the time records of $\overline{\mathcal{J}}$ do not collapse onto a common trend when time is nondimensionalised, indicating that the overall transport intensity is controlled primarily by the velocity magnitude and buoyancy forcing, rather than by the time scaling alone.}

\subsubsection{Experimental field-level transport pathways at fixed nondimensional time}
\label{sec:maps_tstar}

\begin{sidewaysfigure}
\centering
\vspace*{14cm}
\begin{minipage}{\textheight}
\centering
\includegraphics[width=\linewidth]{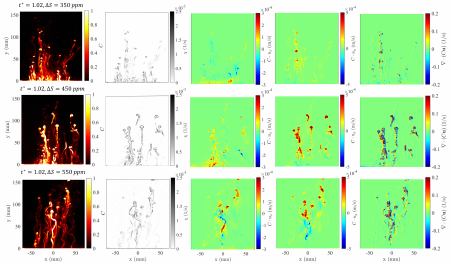}
\captionsetup{width=\linewidth}
\caption{The experimental concentration field $C$ ($1^{\mathrm{st}}$ column), scalar dissipation rate $\chi$ ($2^{\mathrm{nd}}$ column), horizontal \rev{scalar flux} $C u_x$ ($3^{\mathrm{rd}}$ column), vertical \rev{scalar flux} $C u_y$ ($4^{\mathrm{th}}$ column) and divergence of \rev{scalar flux} $\bnabla\!\bcdot (C\boldsymbol{u})$ ($5^{\mathrm{th}}$ column). Data are shown for nondimensional time $t^\ast = 1.02$ for three salinity contrasts: $\Delta S = 350$~ppm (top row), 450~ppm (middle row), and 550~ppm (bottom row).}
\label{fig:maps_tstar102}
\end{minipage}
\end{sidewaysfigure}

To connect the integral measures in figure~\ref{fig:mix_transport} to the underlying local mechanisms, field-resolved diagnostics are compared at the same nondimensional time. Figure~\ref{fig:maps_tstar102} shows the concentration field $C$, scalar dissipation rate $\chi$, horizontal and vertical \rev{scalar flux fields} $C u_x$ and $C u_y$, and the planar divergence $\bnabla\!\bcdot(C\boldsymbol{u})$ at $t^{*}=1.02$ for each salinity contrast.


Clear differences in finger morphology are observed in the concentration fields. For $\Delta S=350$~ppm, broader and more diffuse finger cores are evident, whereas for $\Delta S=550$~ppm, thinner, more intense fingers with sharper scalar gradients are obtained. The corresponding dissipation rate fields $\chi$ indicate that the most intense mixing is concentrated along the finger edges and near the fingertip structures, \rev{with increasingly localised, higher-amplitude dissipation-rate regions as $\Delta S$ increases}, consistent with the formation of sharper scalar gradients under stronger buoyancy forcing.

The \rev{scalar flux fields} provide direct visual evidence of the changing partition between vertical transport and lateral redistribution. The vertical flux $C u_y$ increases in magnitude and becomes more spatially intermittent as $\Delta S$ increases, consistent with the increase in the area-averaged transport in figure~\ref{fig:mix_transport}(c). In parallel, the horizontal flux $C u_x$ becomes increasingly structured and asymmetric at larger $\Delta S$, indicating enhanced lateral transport associated with finger deflection and lateral excursions (zig-zag finger trajectories) during the interaction-dominated stage. The divergence field $\bnabla\!\bcdot(C\boldsymbol{u})$ highlights regions of net convergence/divergence of \rev{scalar transport} within the measurement plane and becomes more patchy and higher-amplitude with increasing $\Delta S$, consistent with stronger local entrainment/detrainment signatures and more finger--finger intermittent interactions.

\subsubsection{Computational three-dimensional flux structure at fixed nondimensional time}
\label{sec:dns_flux}

\begin{figure}
  \centerline{\includegraphics[width=\textwidth]{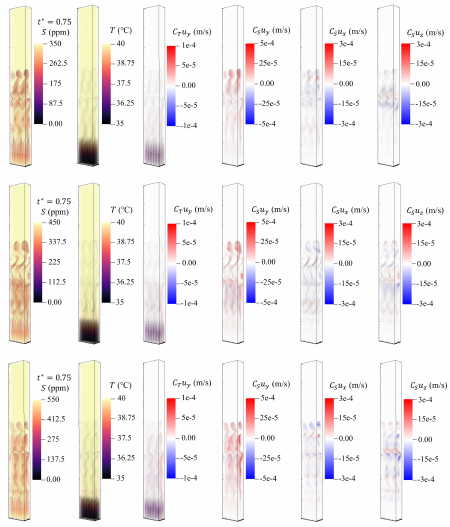}}
  \caption{Comparison of the computational scalar fields and flux components at the same nondimensional time $t^* = 0.75$. Rows correspond to salinity contrast: $\Delta S = 350$ ppm (top row), 450 ppm (middle row), and 550~ppm (bottom row). From left to right, the columns show three-dimensional volume renderings of the salinity field $S$, the temperature field $T$, the vertical \rev{thermal-scalar flux} $C_T u_y$, and the \rev{solutal-scalar flux components} in the vertical ($C_S u_y$), horizontal ($C_S u_x$), and out-of-plane ($C_S u_z$) directions.}
  \label{fig:fluxes}
\end{figure}

The numerical simulations provide access to the full three-dimensional scalar fields and to all components of the \rev{thermal- and solutal-scalar flux vectors}. Figure~\ref{fig:fluxes} presents volume renderings at a fixed nondimensional time $t^{*}=0.75$ for all three salinity contrasts. Shown are the salinity field $S$, temperature field $T$, the vertical \rev{thermal-scalar flux} $C_T u_y$, and the \rev{solutal-scalar flux components} $C_S u_y$, $C_S u_x$, and $C_S u_z$.

Several key features are observed. First, the temperature field exhibits substantially smoother vertical structure than the salinity field, with significant variation confined to a narrow region near the layer interface. This behaviour reflects the large diffusivity contrast (\rev{$Le \approx 75$}), whereby thermal diffusion acts approximately two orders of magnitude faster than salt diffusion. As a consequence, temperature gradients are rapidly smoothed, and regions of strong \rev{thermal-scalar flux} are concentrated close to the layer interface rather than extending throughout the finger length. In contrast, the salinity field maintains sharp, vertically-coherent structures over a much larger portion of the domain.

Second, the magnitude of the vertical \rev{solutal- and thermal-scalar fluxes} increases systematically with increasing $\Delta S$. Although the imposed thermal contrast is fixed across cases, stronger salinity forcing produces more vigorous velocity fields, which in turn amplify both \rev{solutal- and thermal-scalar flux magnitudes} in the vertical direction. Thus, the temperature transport is dynamically enhanced by the salinity-driven buoyancy, even though the thermal gradient itself is unchanged.

Third, at this early-to-intermediate stage ($t^{*}=0.75$), the vertical \rev{thermal-scalar flux} is concentrated very near the interface and is locally comparable to, or larger than, the vertical \rev{scalar flux} in that region. This observation suggests that thermal buoyancy effects play a dominant role during the adjustment phase immediately after initiation, when sharp temperature gradients coexist with rapidly developing velocity disturbances. As the fingers elongate, however, the vertically coherent salinity structures become the primary contributors to sustained upward transport.

Finally, elevated horizontal ($C_S u_x$) and out-of-plane ($C_S u_z$) \rev{solutal-scalar flux components} are observed, particularly for $\Delta S=450$ and $550$~ppm. The presence of these components indicates that \rev{solutal-scalar transport} is intrinsically three-dimensional, with lateral deflection and out-of-plane motion contributing substantially to the total scalar flux. Out-of-plane transport is not directly accessible based on planar measurements, and the results highlight the value of the fully three-dimensional simulations. \rev{Because these quantities are volume-averaged over the reduced periodic domain, their absolute magnitudes and component-wise partition may depend on the domain geometry. The systematic variation with $\Delta S$ is therefore emphasized here.}


\subsubsection{Computational evolution of volume-averaged transport}
\label{sec:dns_flux_evolution}

\begin{figure}
  \centerline{\includegraphics[width=0.9\textwidth]{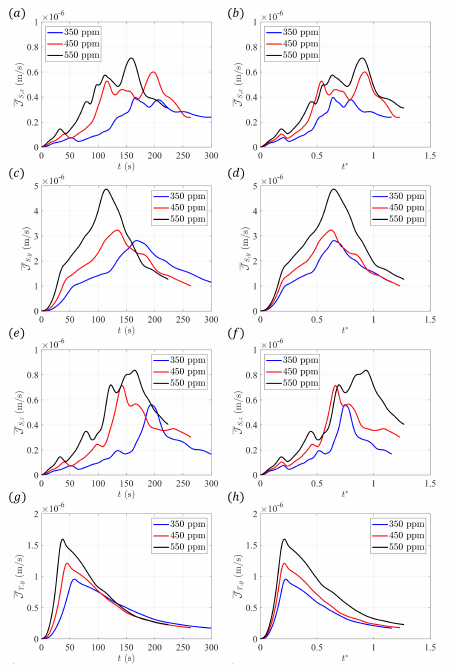}}
  \caption{Temporal evolution of the volume-averaged scalar-flux components from the simulations for different salinity contrasts: (\textit{a},\textit{b}) horizontal \rev{solutal-scalar flux}, $\overline{\mathcal{J}}_{S,x}$; (\textit{c},\textit{d}) vertical \rev{solutal-scalar flux}, $\overline{\mathcal{J}}_{S,y}$; (\textit{e},\textit{f}) out-of-plane \rev{solutal-scalar flux}, $\overline{\mathcal{J}}_{S,z}$; and (\textit{g},\textit{h}) vertical \rev{thermal-scalar flux}, $\overline{\mathcal{J}}_{T,y}$. Results are shown for $\Delta S = 350$, 450, and 550~ppm.}
  \label{fig:fig21_Comp_flux}
\end{figure}

The full temporal evolution of the volume-averaged scalar-flux components is shown in figure~\ref{fig:fig21_Comp_flux} for $\Delta S=350$, 450, and 550~ppm, plotted versus time $t$ and nondimensional time $t^{*}$. In all cases, a transient sequence is obtained in which fluxes rise from near-zero values, reach a well-defined maximum during the most vigorous finger-growth interval, and subsequently decay as the scalar contrasts are depleted and finger motions become increasingly constrained by the upper boundary.

The vertical \rev{solutal-scalar flux component} $\overline{\mathcal{J}}_{S,y}$ (figure~\ref{fig:fig21_Comp_flux}c,d) is found to dominate the \rev{solutal-scalar transport budget} throughout the evolution. A strong increase of its peak magnitude with $\Delta S$ is observed. Peak values of approximately $2.7\times10^{-6}$, $3.3\times10^{-6}$ and $4.9\times10^{-6}$~m~s$^{-1}$ are obtained for $\Delta S=350$, 450, and 550~ppm, respectively. In dimensional time, the time-to-peak decreases systematically with increasing $\Delta S$ (the maximum being reached at earlier $t$ for larger $\Delta S$), consistent with the faster finger growth and earlier interaction with the upper boundary under stronger haline buoyancy forcing. The peaks occur over a narrow nondimensional time interval (approximately $t^{*}\approx0.6$--$0.8$), indicating that the \emph{timing} of the strongest vertical transport is largely set by the same growth-time scaling used to collapse the finger-height histories, whereas the \emph{magnitude} increases with $\Delta S$.

The lateral \rev{solutal-scalar flux components} $\overline{\mathcal{J}}_{S,x}$ and $\overline{\mathcal{J}}_{S,z}$ (figure~\ref{fig:fig21_Comp_flux}a,b,e,f) remain smaller than $\overline{\mathcal{J}}_{S,y}$, but a systematic strengthening with $\Delta S$ is also observed. \rev{Multiple local peaks are visible} in the time \rev{histories of} $\overline{\mathcal{J}}_{S,x}$ and $\overline{\mathcal{J}}_{S,z}$, particularly at the higher salinity contrasts, indicating that lateral transport is modulated intermittently during the interaction stage rather than evolving as a single smooth pulse. This behaviour is consistent with episodic lateral deflection and three-dimensional reorganisation of fingers, as suggested by the volume-rendering snapshots in figure~\ref{fig:fluxes}.



The vertical \rev{thermal-scalar flux} $\overline{\mathcal{J}}_{T,y}$ follows a qualitatively different temporal evolution (figure~\ref{fig:fig21_Comp_flux}g,h). A sharp early-time peak is obtained for all cases, followed by a comparatively long decay. Peak values of approximately $1.0\times10^{-6}$, $1.2\times10^{-6}$ and $1.6\times10^{-6}$~m~s$^{-1}$ are obtained for $\Delta S=350$, 450, and 550~ppm, respectively, and the peak occurs much earlier than the peak in $\overline{\mathcal{J}}_{S,y}$. In nondimensional time, the \rev{thermal-scalar-flux} maximum occurs near the beginning of the evolution (approximately $t^{*}\approx0.2$), whereas the strongest vertical \rev{solutal-scalar transport} occurs later at $t^{*}\approx0.65$--$0.75$. This separation in time is consistent with the large diffusivity contrast (\rev{$Le\approx75$}): rapid thermal diffusion \rev{localises} strong temperature gradients near the interface and produces an early burst of vertical temperature transport, whereas the \rev{solutal-scalar flux} builds more slowly as vertically coherent fingers develop and intensify.

Overall, the computational trends are consistent with the experimental observations. The monotonic increase of the peak vertical \rev{solutal-scalar transport magnitude} with $\Delta S$ in the DNS \rev{parallels} the experimentally measured increase in the \rev{area-averaged scalar transport magnitude} $\overline{\mathcal{J}}$ (figure~\ref{fig:mix_transport}), \rev{showing consistently that stronger haline buoyancy intensifies advective scalar transport}. The earlier time-to-peak at larger $\Delta S$ is likewise consistent with the experimentally observed acceleration of finger growth and earlier onset of rapid mixing. The strengthening of the horizontal DNS \rev{solutal-scalar transport magnitudes} also corresponds to the experimentally observed increase in lateral finger deflection, while the early \rev{thermal-scalar transport} pulse reflects rapid thermal adjustment near the interface.


\subsection{Finger-scale dynamics: Symmetric vortex-ring regime ($\Delta S = 450$ ppm)}
\label{sec:finger-scale}

Finger-scale dynamics are examined using a representative salt finger from the $\Delta S=450$~ppm salinity-contrast case. This case is chosen as an intermediate set of conditions between relatively weak forcing ($\Delta S=350$~ppm), for which growth and transport are slower, and stronger forcing ($\Delta S=550$~ppm), for which pronounced lateral excursions and stronger three-dimensionality are observed (addressed in \S\ref{sec:zigzag}, below). The representative finger \rev{analysed} here is the same structure shown previously in figure~\ref{fig:registry}(c), in which the fidelity of the PIV--PLIF registration and the mean-subtracted velocity field were demonstrated.

For clarity, all fields in figures~\ref{fig:singlefinger_450}--\ref{fig:timeseries_450} are reported for a fixed zoom window enclosing the finger, and the displayed velocity vectors correspond to the \emph{mean-subtracted} velocity,

\begin{equation}
\boldsymbol{u}'(x,y,t) \equiv \boldsymbol{u}(x,y,t)-\overline{\boldsymbol{u}}(t),
\label{eq:uprime_def}
\end{equation}
where $\overline{\boldsymbol{u}}(t)$ is the area-averaged velocity within the zoomed-in window at each time. Equation \ref{eq:uprime_def} removes the bulk advection and isolates the relative motion associated with the coherent finger circulation.

\subsubsection{Vortex-ring formation and mushroom-tip evolution}
\label{sec:vortexpair_450}

\begin{sidewaysfigure}
\centering
\vspace*{14cm}
\begin{minipage}{\textheight}
\centering
\includegraphics[width=0.9\linewidth]{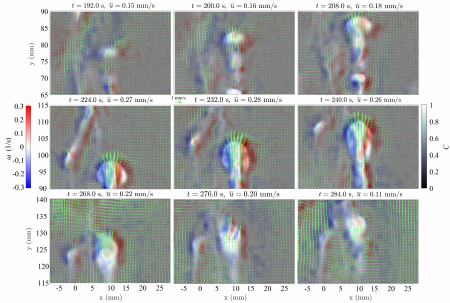}
\captionsetup{width=\linewidth}
\caption{Temporal evolution of a single salt finger for the $\Delta S = 450$~ppm experimental case, tracked from its initial formation through growth, deformation, and eventual breakdown and mixing. Each panel shows mean-subtracted velocity vectors (green) overlaid on the vorticity field (red--blue colormap) and the concentration field (grayscale colormap). The time point and the subtracted average velocity are indicated above each panel.}
\label{fig:singlefinger_450}
\end{minipage}
\end{sidewaysfigure}

The temporal evolution of the selected finger is shown in figure~\ref{fig:singlefinger_450}, where the velocity vectors are overlaid on the vorticity field and the \rev{normalised solutal scalar field}. Coherent counter-rotating vortex cores are observed to develop on either side of the finger stem\rev{. These cores correspond to the intersection of the measurement plane with a vortex ring and reveal} the characteristic mushroom-shaped tip. The evolution may be described in three stages.


During the early formation stage (top row), a small bulb forms at the advancing tip while a narrow neck is observed beneath it. Thin shear layers appear along the stem and near the forming cap, and the vorticity distribution remains relatively compact. The mean upward velocity within the zoomed-in region remains relatively small during this stage ($\bar{u}\approx0.15$--$0.18$~mm~s$^{-1}$), confirming that the finger is still in an acceleration phase. Upward propagation is dominant, and lateral displacement remains limited.

During the intermediate steady-elongation stage (middle row), finger growth proceeds with an approximately constant upward propagation rate, consistent with the intermediate plateau in the ensemble growth-rate curves (figure~\ref{fig:growth}c). The mean velocity increases substantially and reaches its maximum ($\bar{u}\approx0.27$--$0.28$~mm~s$^{-1}$), indicating the period of strongest vertical momentum. The mushroom cap enlarges, the neck stretches, and the vorticity layers surrounding the tip broaden. The counter-rotating vortex cores are strongest in this stage, indicating enhanced momentum exchange and persistent interfacial entrainment and scalar transport.

During the late stage (bottom row), coherence of the tip is progressively reduced. The cap deforms, the neck structure breaks down, and weakening of the vortex cores is observed together with increased local mixing and loss of sharp interfacial structure. The mean velocity correspondingly decreases ($\bar{u}\approx0.20$~mm~s$^{-1}$ and eventually $\approx0.11$~mm~s$^{-1}$), reflecting the decay of organised upward motion. These features indicate the onset of finger breakdown and transition toward a more mixed state.

\subsubsection{Local mixing and vertical transport within a single finger}
\label{sec:chi_flux_450}

\begin{figure}
  \centerline{\includegraphics[width=\textwidth]{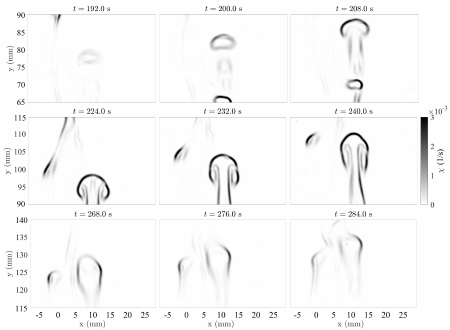}}
  \caption{Temporal evolution of the scalar dissipation rate for a representative salt finger for the $\Delta S = 450$~ppm experimental case (the same salt finger shown in figure~\ref{fig:singlefinger_450}), covering the finger’s development from initial formation to breakdown and mixing.}
  \label{fig:chi_450}
\end{figure}

\begin{figure}
  \centerline{\includegraphics[width=\textwidth]{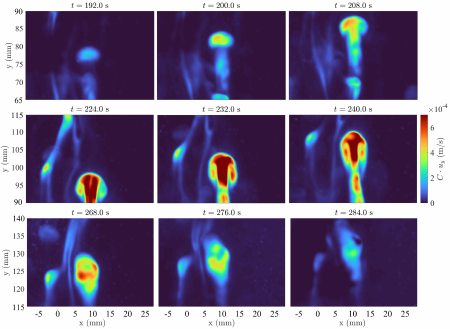}}
  \caption{Temporal evolution of the vertical \rev{scalar flux field}, $C u_y$, for a representative salt finger for the $\Delta S = 450$~ppm experimental case (the same salt finger shown in figure~\ref{fig:singlefinger_450}), covering its development from initial formation through mature growth and eventual breakdown and mixing.}
  \label{fig:fluxy_450}
\end{figure}

The coupling between local mixing and convective transport is quantified using the scalar dissipation rate and the vertical scalar-flux field for the same finger. A time sequence of the $\chi$ and $C u_y$ fields is shown in figures~\ref{fig:chi_450} and \ref{fig:fluxy_450}, respectively, over the same time interval as figure~\ref{fig:singlefinger_450}.

In the early stage (top row), regions of elevated scalar dissipation rate are confined to a compact region near the forming tip, reflecting the initial sharpening of scalar gradients as the cap emerges and the neck forms. A \rev{localised} region of weakly elevated vertical flux is observed at the same location, indicating the onset of convective scalar transport.

In the intermediate stage (middle row), the elevated regions of $\chi$ and $C u_y$ intensify substantially. The dissipation rate field becomes strongest around the mushroom-shaped cap and along the edges of the stem, consistent with the locations of strong straining and roll-up that sharpen the scalar interfaces. Simultaneously, the vertical-flux field develops a pronounced jet-like core in the stem beneath the cap. In addition, side-lobe regions of elevated $Cu_y$ appear adjacent to the core, consistent with regions where surrounding fluid is entrained into the finger during the period of strongest transport.

In the late stage (bottom row), the strength and extent of the elevated regions of $\chi$ and $C u_y$ decrease as the finger deforms and loses coherence. Scalar gradients are smoothed, the dissipation rate field becomes weaker and more diffuse, and the vertical-flux field becomes less symmetric and more intermittent. This behaviour marks the transition from organised finger-driven transport to a mixed, diffusion-dominated regime.


\subsubsection{Time evolution of circulation, enstrophy, dissipation rate, and flux}
\label{sec:timeseries_450}

\begin{figure}
  \centerline{\includegraphics[width=\textwidth]{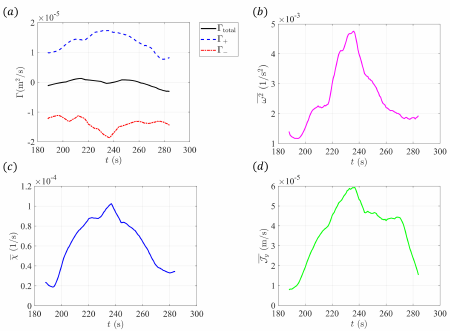}}
  \caption{Temporal evolution of integrated quantities for a representative salt finger for the $\Delta S = 450$~ppm experimental case, showing its evolution from formation to breakdown: (\textit{a}) circulation for the counter-clockwise rotating fluid (blue dashed line) and clockwise rotating fluid (red dash-dotted line) and the total circulation (black line). (\textit{b}) Area-averaged enstrophy. (\textit{c}) Area-averaged scalar dissipation rate. (\textit{d}) Area-averaged vertical \rev{scalar flux}. All quantities are computed for the zoomed-in field of view surrounding the finger and span the same time period shown in figures~\ref{fig:singlefinger_450}--\ref{fig:fluxy_450}.}
  \label{fig:timeseries_450}
\end{figure}

The qualitative vortex-ring dynamics \rev{visualised} in figures~\ref{fig:singlefinger_450}--\ref{fig:fluxy_450} are quantified using integrated measures computed over the same region enclosing the representative finger. The temporal evolution of these area-integrated quantities is shown in figure~\ref{fig:timeseries_450}.

The circulation is defined as the area integral of vorticity within the zoom window,

\begin{equation}
\Gamma(t) \equiv \int_A \omega(x,y,t)\,\mathrm{d}A.
\label{eq:Gamma_def}
\end{equation}
To distinguish the two vortex cores, the circulation is separated into counter-clockwise and clockwise contributions, $\Gamma^{+}$ and $\Gamma^{-}$, as well as their sum $\Gamma_{\mathrm{total}}=\Gamma^{+}+\Gamma^{-}$ (figure~\ref{fig:timeseries_450}a).

Approximately equal and opposite values of $\Gamma^{+}$ and $\Gamma^{-}$ are obtained throughout the interval, hence $\Gamma_{\mathrm{total}}$ remains close to zero. This behaviour quantitatively confirms the formation of a nearly symmetric \rev{pair of} counter-rotating vortex cores that roll up the scalar interface to produce the mushroom-shaped cap observed in figure~\ref{fig:singlefinger_450}.

The area-averaged enstrophy, scalar dissipation rate, and vertical scalar flux (figures~\ref{fig:timeseries_450}b–d) rise during the growth stage and reach maxima at roughly the same time point ($\sim240$ s), corresponding to the period of strongest vortex-ring intensity and maximum mean upward velocity. The concurrent peaks indicate that intensified vorticity production sharpens the concentration gradients, increasing the dissipation rate, while simultaneously enhancing upward scalar transport. Following the peak, all three quantities decrease, indicating weakening of the vortex cap, gradient smoothing of the scalar field, and decay of convective transport as the finger transitions toward breakdown and mixing with the surrounding fluid.

\subsection{Finger-scale dynamics: zig-zag and lateral-drift regime ($\Delta S = 550$ ppm)}
\label{sec:zigzag}

Increasing the salinity contrast to $\Delta S=550$~ppm (lower density ratio and stronger solutal buoyancy forcing) produces a qualitative change in finger-scale dynamics. Unlike the predominantly vertical, symmetric growth observed at $\Delta S=450$~ppm, the fingers now exhibit intermittent, pronounced lateral drift with alternating deflection directions.

\begin{figure}
  \centerline{\includegraphics[width=\textwidth]{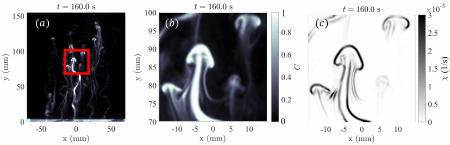}}
  \caption{Example of a representative salt finger for the $\Delta S = 550$~ppm experimental case at $t = 160$~s. (\textit{a}) Full concentration field showing the finger formation, with the red box highlighting the zoom region. (\textit{b}) Concentration field for the zoomed-in region, capturing the head of the finger structure. (\textit{c}) Scalar dissipation rate field, $\chi$, for the zoomed-in region.}
  \label{fig:example_550}
\end{figure}

Figure~\ref{fig:example_550} shows a representative finger selected for detailed analysis. The full-field view (figure~\ref{fig:example_550}a) demonstrates the lateral excursions of the finger trajectories. The zoomed-in concentration field (figure~\ref{fig:example_550}b) shows a tilted, asymmetric cap, while the corresponding scalar dissipation rate field (figure~\ref{fig:example_550}c) reveals that the dissipation rate is no longer symmetrically distributed about the tip. Instead, the regions of elevated dissipation rate concentrate preferentially on one side, consistent with lateral shear and asymmetric vortex development.

\begin{figure}
  \centerline{\includegraphics[width=\textwidth]{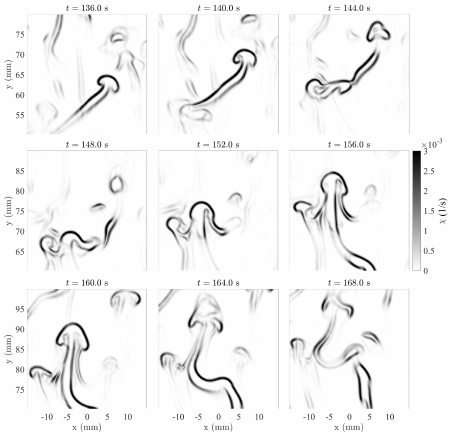}}
  \caption{Temporal evolution of the scalar dissipation rate field, $\chi$, for representative salt fingers for the $\Delta S = 550$~ppm experimental case.}
  \label{fig:chi_sequence_550}
\end{figure}

A time sequence of the scalar dissipation rate field (figure~\ref{fig:chi_sequence_550}) illustrates the emergence of zig-zag finger trajectories. In the early stage (top row), a finger forms and rapidly elongates. However, instead of maintaining vertical alignment, the tip deflects laterally under strong buoyancy forcing.

A subsequent finger emerges below the first finger and deflects in the opposite direction (middle row). This alternating \rev{behaviour} continues over time (bottom row), producing a series of zig-zag trajectories. The scalar dissipation rate field highlights thin, high-gradient layers that wrap asymmetrically around the deforming tip, revealing inherently asymmetric roll-up dynamics. Further, this \rev{behaviour} contrasts sharply with the $\Delta S=450$~ppm case (discussed in \rev{\S\ref{sec:timeseries_450}}), in which the vortex ring remains nearly symmetric and \rev{vertical} propagation dominates.


\begin{figure}
  \centerline{\includegraphics[width=\textwidth]{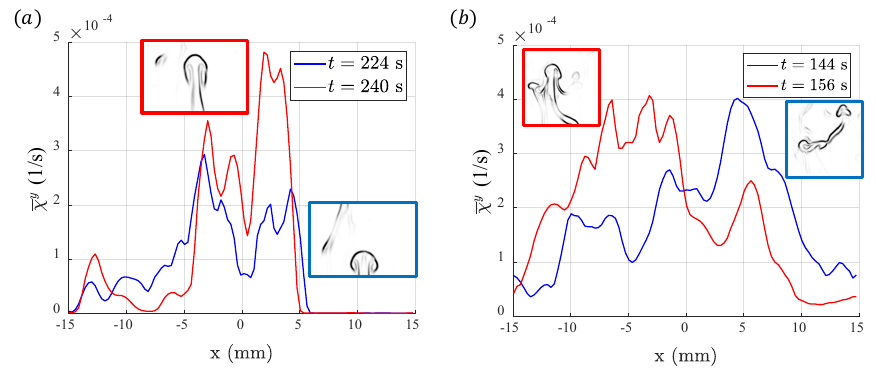}}
  \caption{Vertically-averaged profiles of scalar dissipation rate $\overline{\chi}^{\,y}$ at two time points for a representative finger for (\textit{a}) $\Delta S = 450$~ppm and (\textit{b}) $\Delta S = 550$~ppm experimental cases. The two profiles in each panel are separated by the same nondimensional time interval, $\Delta t^* = 0.074$. Insets show the corresponding $\chi$ fields at the same time points (the field with the red border corresponds to the red profile; the field with the blue border corresponds to the blue profile).}
  \label{fig:chi_y_profiles}
\end{figure}

The asymmetric \rev{deflections} are quantified in figure~\ref{fig:chi_y_profiles}, which compares vertically-averaged dissipation rate profiles at matched nondimensional times for the $\Delta S=450$ and 550~ppm cases. For $\Delta S=450$~ppm (figure~\ref{fig:chi_y_profiles}a), the dissipation rate profile exhibits a single, centrally-located elevated region corresponding to the symmetric mushroom cap. In contrast, for $\Delta S=550$~ppm (figure~\ref{fig:chi_y_profiles}b), the elevated region in the profile is offset from the centerline, to the left at $t=144$~s and to the right at $t=156$~s. These offset elevated regions reflect the sideways displacement of the tip and the formation of asymmetric shear layers along the sides of the finger. The asymmetric dissipation rate profiles provide quantitative evidence that the scalar interface is oriented obliquely rather than vertically upward.



\begin{figure}
  \centerline{\includegraphics[width=\textwidth]{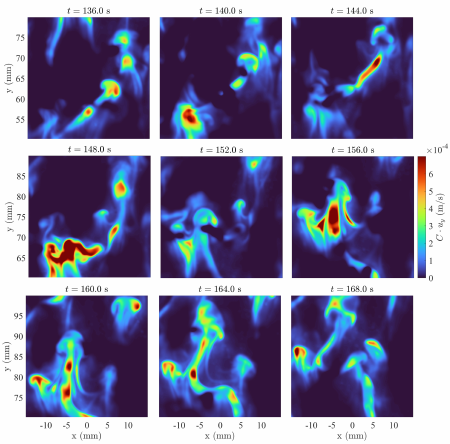}}
  \caption{Temporal evolution of the vertical \rev{scalar flux field}, $C u_y$, for representative salt fingers for the $\Delta S = 550$~ppm experimental case (same time sequence as shown in figure~\ref{fig:chi_sequence_550}).}
  \label{fig:fluxy_sequence_550}
\end{figure}

The transport fields provide further insight into the finger growth, evolution, and dynamics. The sequence of vertical scalar-flux fields (figure~\ref{fig:fluxy_sequence_550}) shows narrow jet-like cores beneath each evolving tip, similar to the lower-salinity case. However, these vertical flux cores now appear tilted with significant horizontal displacement.

\begin{figure}
  \centerline{\includegraphics[width=\textwidth]{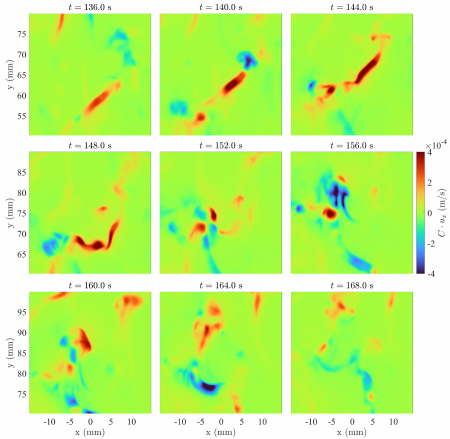}}
  \caption{Temporal evolution of the horizontal \rev{scalar flux field}, $C u_x$, for representative salt fingers for the $\Delta S = 550$~ppm experimental case (same time sequence as shown in figure~\ref{fig:chi_sequence_550}). The sequence shows the fingers' development from the early stage through growth, deformation, and eventual breakdown.}
  \label{fig:fluxx_sequence_550}
\end{figure}

\rev{Consistently}, the sequence of horizontal flux fields (figure~\ref{fig:fluxx_sequence_550}) reveals elevated regions of alternating positive and negative $C u_x$ around the fingers. As each finger tilts, horizontal entrainment zones form on opposing sides of the stem, generating alternating patches of \rev{leftward and rightward} horizontal scalar transport. The vertical and horizontal flux fields are clearly coupled and collectively quantify scalar transport along the tilted finger trajectories, further revealing that transport is no longer dominated by vertical convection alone in the higher-salinity-contrast case.




\rev{A qualitatively similar transition toward zig-zag and lateral-drift behaviour occurs in both the experimental PLIF/PIV fields and the DNS as $\Delta S$ increases, despite the different lateral boundary conditions and form of the initial disturbances. The DNS also shows a corresponding strengthening of the horizontal and out-of-plane solutal-scalar transport components, $\overline{\mathcal{J}}_{S,x}$ and $\overline{\mathcal{J}}_{S,z}$ (figures~\ref{fig:fluxes} and~\ref{fig:fig21_Comp_flux}). This cross-configuration agreement supports the interpretation that the transition is a physical response to stronger double-diffusive forcing, although its precise onset and quantitative flux partition may remain sensitive to the domain geometry.}

\rev{Taken together, the experimental and numerical observations indicate that the zig-zag regime involves a loss of symmetry of the coherent circulation around the fingertip as the double-diffusive forcing increases. Lateral and sinuous motions have also been reported in conventional buoyant plumes, including the buckling of thermal plumes and global sinuous instabilities of laminar plumes in stratified environments \citep{yang1992buckling,lombardi2011growth}. In those source-driven flows, the deformation develops along an extended plume structure, whereas the behaviour observed here is localised around the transient mushroom-shaped tip and stem of individual double-diffusive fingers. The present transition also shares features with shear-driven secondary instabilities previously reported in salt-finger flows \citep{lambert1972vertical,linden1973structure,taylor1996experiments,pringle2002double} and with symmetry breaking of vortex-ring structures, for which loss of axisymmetry can produce oblique trajectories \citep{webster1998vortex}. Because increasing $\Delta S$ simultaneously lowers the density ratio and changes the solutal Rayleigh number, fingertip velocity, and shear along the finger stem, the present cases do not isolate a unique critical Reynolds- or Rayleigh-number threshold. The combined dissipation-rate and scalar-flux evidence is therefore consistent with a shear-mediated loss of vortex-ring symmetry under stronger double-diffusive forcing, without assigning the observed behaviour to a unique classical instability mode.}

\subsection{Buoyancy perturbation dynamics and force balance governing finger growth}
\label{sec:buoyancy_dynamics}


To build on the preceding analyses that established the phenomenology of finger growth, transport, and shear-layer development, the underlying buoyancy forcing responsible for these dynamics is examined directly using the three-dimensional simulations. Particular emphasis is placed on how the competing thermal and solutal contributions evolve in time and regulate the observed three-stage growth-rate \rev{behaviour}.


Under the Boussinesq approximation, buoyancy is defined as


\begin{equation}
b = g\left[\alpha (T - T_{\mathrm{ref}}) - \beta (S - S_{\mathrm{ref}})\right],
\end{equation}

\noindent
where $T_{\mathrm{ref}}=35^\circ$C is the cold reference temperature, and $S_{\mathrm{ref}}=0$ ppm is the fresh reference state. To isolate the dynamically active component, the horizontally averaged buoyancy profile is removed,
\begin{equation}
b'(x,y,z,t) = b(x,y,z,t) - \overline{b}^{\,x,z}(y,t),
\end{equation}

\noindent
where the overbar denotes averaging in the horizontal $(x,z)$ directions for a fixed value of $y$. The perturbation field $b'$ therefore represents buoyancy anomalies associated with the finger structures, whereas the mean profile $\overline{b}^{\,x,z}$ maintains the background stratification.

For interpretation, the buoyancy anomaly may be written as the sum of thermal and solutal contributions,

\begin{equation}
b' = b_T' + b_S',
\qquad
b_T' \equiv g\alpha\left(T-\overline{T}^{\,x,z}\right),
\qquad
b_S' \equiv -g\beta\left(S-\overline{S}^{\,x,z}\right).
\label{eq:b_decomp}
\end{equation}

\par
\begin{figure}
  \centerline{\includegraphics[width=\textwidth]{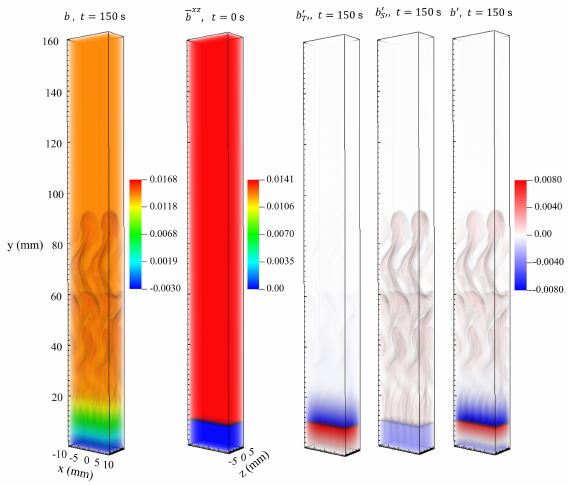}}
  \caption{Three-dimensional volume renderings of the buoyancy $b$ and its decomposition at representative times in the simulations. From left to right: the total buoyancy field $b$ at $t=150~\mathrm{s}$ for the $\Delta S = 450$~ppm case; the horizontally averaged ($x$–$z$) buoyancy field $\overline{b}^{xz}$ at $t=0~\mathrm{s}$; the temperature-dominated buoyancy anomaly $b_T'$ at $t=150~\mathrm{s}$; the salinity-dominated buoyancy anomaly $b_S'$ at $t=150~\mathrm{s}$; and the total buoyancy anomaly $b'$ at $t=150~\mathrm{s}$.}
  \label{fig:buoyancy}
\end{figure}

\noindent
In the present configuration, negative buoyancy anomalies are predominantly associated with the thermal contribution (i.e., due to the rapidly diffusing temperature field), whereas positive buoyancy anomalies are predominantly associated with the salinity contribution (i.e., due to the slowly diffusing salt field). These competing effects are illustrated in figure~\ref{fig:buoyancy}, in which regions of elevated $b_T'$ remain confined to a thin region near the interface, whereas the $b_S'$ field exhibits vertically coherent anomalies associated with the developing fingers.

Vertical fluid motion is governed by the $y$-momentum equation,
\begin{equation}
\frac{\partial u_y}{\partial t}
+ \boldsymbol{u}\cdot\nabla u_y
= -\frac{1}{\rho_0}\frac{\partial p'}{\partial y}
+ b'
+ \nu \nabla^2 u_y.
\label{eq:vertical_momentum}
\end{equation}
Here, \rev{$\rho_0$ is the reference density, $p'$ is the pressure perturbation, and} $b'$ acts as a body force per unit mass. The fingertip motion and propagation speed are governed by the balance of $b'$, pressure-gradient, and shear forces as shown in equation \eqref{eq:vertical_momentum}. The observed transition \rev{from acceleration to approximately constant propagation speed and then deceleration is consistent with} a corresponding transition from increasing net forcing to approximate force balance \rev{and subsequently} to weakening net forcing.

\begin{figure}
  \centerline{\includegraphics[width=\textwidth]{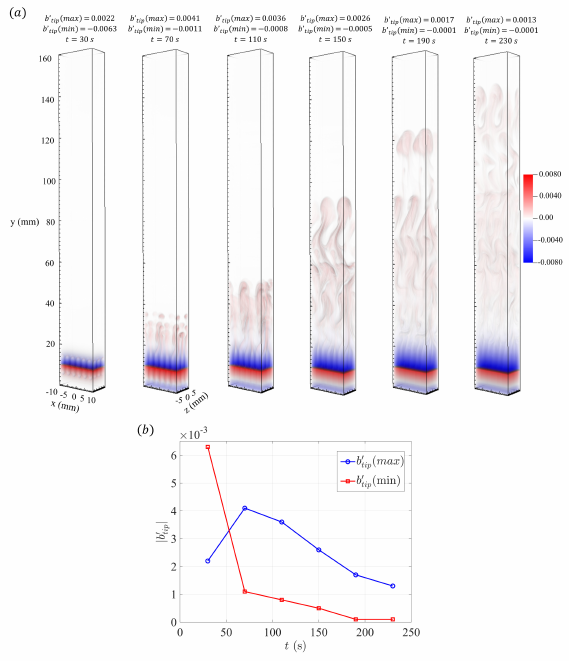}}
  \caption{(\textit{a}) Three-dimensional volume renderings of the buoyancy anomaly $b'$ at selected time points during the development of finger-type convection for the $\Delta S = 450$~ppm case in the simulation, with the maximum and minimum of the fingertip buoyancy anomaly $b'_{\mathrm{tip}}$ indicated for each time. (\textit{b}) Temporal evolution of the corresponding extrema, showing $b'_{\mathrm{tip}}(\max)$ and $-\,b'_{\mathrm{tip}}(\min)$ as functions of time.}
  \label{fig:bPrime}
\end{figure}

The temporal evolution of the fingertip buoyancy anomaly is summarised in figure~\ref{fig:bPrime} for the $\Delta S=450$~ppm case. Figure~\ref{fig:bPrime}(a) presents the three-dimensional volume renderings of $b'$ at selected times, with the corresponding extrema of $b'_{\mathrm{tip}}$ listed above each rendering, while figure~\ref{fig:bPrime}(b) shows the temporal evolution of the $b'_{\mathrm{tip}}$ extrema. The maximum, $b'_{\mathrm{tip}}(\max)$, is associated with the salinity-dominated positive buoyancy anomaly at the rising fingertip. The minimum, $b'_{\mathrm{tip}}(\min)$, is associated with the temperature-dominated negative buoyancy anomaly that persists within the near-tip region owing to rapid thermal adjustment and entrainment. A clear temporal asymmetry between these two extrema is observed, as seen most directly in figure~\ref{fig:bPrime}(b); at early times the magnitude of the negative anomaly exceeds that of the positive anomaly, but the negative anomaly decays rapidly, whereas the positive anomaly persists longer and ultimately governs the sustained finger rise.

Quantitatively, the early-time values reported in figure~\ref{fig:bPrime} show that at $t=30$~s, $b'_{\mathrm{tip}}(\min)\approx-0.0063$ while $b'_{\mathrm{tip}}(\max)\approx0.0022$. By $t=70$~s, the negative anomaly has weakened substantially to $b'_{\mathrm{tip}}(\min)\approx-0.0011$, while the positive anomaly has increased to $b'_{\mathrm{tip}}(\max)\approx0.0041$. This rapid reduction of the negative (temperature-dominated) anomaly is consistent with fast thermal diffusion ($Le\gg1$), which rapidly smooths temperature gradients and confines thermal buoyancy anomalies to a thin region near the initial interface. In contrast, the positive (salinity-dominated) anomaly strengthens initially as vertically coherent salt fingers develop while the salinity gradients remain sharp.

After $t\approx70$~s, however, $b'_{\mathrm{tip}}(\max)$ decreases gradually from the local maximum: $b'_{\mathrm{tip}}(\max)\approx0.0036$ at $t=110$~s and $\approx0.0026$ at $t=150$~s, with further reduction thereafter. This gradual decline is especially clear in figure~\ref{fig:bPrime}(b), while figure~\ref{fig:bPrime}(a) shows the accompanying weakening and spreading of the positive buoyancy anomaly near the fingertips. This decline coincides with the onset of the intermediate regime of nearly constant fingertip growth rate (figures~\ref{fig:growth} and~\ref{fig:compGrowth}), indicating that, once the fingers are fully developed, the buoyancy anomaly at the tip erodes by entrainment and mixing, even though salinity diffusion remains slow. The temporal evolution of $b'_{\mathrm{tip}}$ provides a direct scalar-based explanation for the transitions in tip propagation, maps onto the three-stage growth-rate \rev{behaviour} identified above, and enables interpretation of the dynamics in terms of a time-dependent force balance:

\emph{(i) Initial acceleration phase.}
During the initial stage, the net buoyancy forcing acting on the developing fingertips strengthens. This is reflected in the rapid growth of $b'_{\mathrm{tip}}(\max)$ between $t=30$~s and $t\approx70$~s (figure~\ref{fig:bPrime}b), while the competing thermal anomaly $b'_{\mathrm{tip}}(\min)$ decays rapidly. The renderings in figure~\ref{fig:bPrime}(a) further show that the temperature-dominated negative anomaly remains concentrated near the lower interfacial region, whereas the salinity-dominated positive anomaly becomes increasingly associated with the upward-growing fingers. The dominance of the positive (salinity-induced) anomaly after this early adjustment period implies an increasing upward body force in equation \eqref{eq:vertical_momentum}, producing accelerating fingertip motion and the increasing growth rate observed both experimentally and computationally.






\emph{(ii) Quasi-steady propagation phase.}
Once $t\gtrsim70$~s, $b'_{\mathrm{tip}}(\max)$ begins to decrease, yet the fingertip \rev{propagation rate} remains approximately constant over an extended interval. This behaviour indicates that the tip dynamics are no longer controlled by the increasing buoyancy anomaly, but rather by an approximate balance between buoyancy forcing and resistance. In this regime, the ascending fingers develop strong shear layers along their boundaries (as documented in \S\ref{sec:finger-scale}), which roll up into a mushroom-shaped head and drive entrainment. The associated vorticity and entrainment increase both the scalar dissipation rate and form drag while simultaneously diluting the buoyancy anomaly at the tip. The combined effect is a self-regulating feedback in which buoyancy forcing and shear-induced resistance remain comparable, yielding \rev{negligible fingertip acceleration and an approximately constant fingertip propagation speed} (figure~\ref{fig:growth}c).

\emph{(iii) Decay and boundary-influenced phase.}
At later times, further reduction of $b'_{\mathrm{tip}}(\max)$ is observed (figure~\ref{fig:bPrime}b), consistent with continued entrainment-driven dilution and progressive depletion of local salinity gradients. The net forcing decreases as the buoyancy anomaly weakens relative to the opposing viscous and pressure-gradient effects, leading to the observed decline in growth rate near the top boundary. Figure~\ref{fig:bPrime}(a) shows that, by these later times, the buoyancy-anomaly field has become weaker and more spatially diffuse in the tip region, consistent with this decay-stage interpretation.

The buoyancy decomposition and its extrema, therefore, provide a mechanistic explanation for the growth-rate evolution sequence. Rapid thermal diffusion causes the temperature-dominated (negative) buoyancy anomaly to decay quickly, allowing salinity-dominated (positive) buoyancy anomalies to dominate and drive the finger rise. The subsequent decay of the salinity-driven anomaly after $t\approx70$~s coincides with the onset of the constant-growth-rate regime, indicating that shear-layer development and entrainment introduce sufficient resistance and dilution to establish an approximate force balance. Finally, continued dilution and boundary influence weaken $b'_{\mathrm{tip}}$ further, producing deceleration and growth-rate decay. Together, these results connect the competing thermal and solutal diffusion pathways, the evolution of buoyancy forcing, and the shear-layer physics observed at the finger scale into a combined, time-dependent force-balance description of finger-type double-diffusive convection.

\section{Conclusions}
\label{sec:conclusions}

The combined experimental and computational dataset provides finger-resolved insights into transient fingering DDI and yields the following conclusions.

(i) For each salinity contrast level, a reproducible three-stage growth history was obtained in the experimental data: an early acceleration phase, an intermediate quasi-steady propagation phase with nearly constant tip speed, and a late decay phase as finger interactions and top-boundary effects intensified. \rev{The measured peak/plateau growth rates increased monotonically with salinity contrast and were closely reproduced by the DNS results (table~\ref{tab:hdot_comparison}). When nondimensionalised, the fingertip-height histories collapsed onto a common trend within the investigated parameter range.}

(ii) Transport magnitude increases strongly with salinity forcing and includes significant non-vertical pathways at low $R_{\rho}$. For $\Delta S=450$~ppm, a symmetric vortex-ring regime was observed, in which coherent counter-rotating cores formed around the stem, rolled up the interface into a mushroom-shaped cap, and produced co-located regions of elevated enstrophy, scalar dissipation rate, and vertical flux within a single finger during the steady-propagation stage. \rev{At $\Delta S=550$~ppm, both the experimental and DNS results exhibited a qualitative transition toward zig-zag/lateral-drift behaviour for the fingers. The occurrence of this behaviour in both approaches supports its interpretation as a robust response to stronger double-diffusive forcing, although its exact onset and spatial organisation may remain sensitive to the domain geometry and boundary conditions.}

(iii) The transition from accelerating to constant-speed to decelerating growth \rev{is consistent with} a buoyancy--shear balance that can be quantified directly from the buoyancy perturbation evolution. In the early stage, the fingertip buoyancy anomaly exhibited a rapid adjustment in which the temperature-dominated negative anomaly weakened quickly, whereas the salinity-dominated positive anomaly strengthened to a maximum, thus combining to produce fingertip acceleration. The constant tip speed in the mid-stage resulted from an approximate force balance between buoyancy forcing and shear-induced resistance, and the late-stage decay was attributed to continued dilution of buoyancy anomalies and increasing influence of the upper boundary.

These results offer validation targets for future modelling and provide mechanistic constraints for reduced descriptions of DDI-driven mixing and transport in stratified environments.

\section*{Acknowledgements}
The authors gratefully acknowledge financial support from the Karen and John Huff Chair endowment and a CPI award from the Georgia Tech Space Research Institute. The support of Ewan Pritchard and Blaire Doss in the experimental work is gratefully acknowledged.

The authors also acknowledge the generous computing resources from the DOE's 2025 ALCC award (TUR147 \& BubbleLaden, PI: Jain). This research used supporting resources at the Argonne and the Oak Ridge Leadership Computing Facilities. The Argonne Leadership Computing Facility at Argonne National Laboratory is supported by the Office of Science of the U.S. DOE under Contract No. DE-AC02-06CH11357. The Oak Ridge Leadership Computing Facility at the Oak Ridge National Laboratory is supported by the Office of Science of the U.S. DOE under Contract No. DE-AC05-00OR22725.





\bibliographystyle{jfm}
\bibliography{DDI}
\end{document}